\documentclass[notitlepage,
twocolumn,superscriptaddress,amsmath,amssymb,aps,pra]{revtex4-2}
\usepackage{amsmath,amssymb,graphicx,mathtools,dsfont,bbold, natbib}
\usepackage[dvipsnames]{xcolor}
\usepackage[T1]{fontenc}
\usepackage{hyperref}
\usepackage{soul}
\usepackage{bibunits}
\usepackage{dcolumn}
\usepackage{bm}
\usepackage{color}
\usepackage{physics}
\usepackage{dsfont}
\usepackage{adjustbox}
\usepackage{soul}
\usepackage{amsmath,amssymb,physics}
\usepackage{multirow}

\usepackage{ulem}
\usepackage{rotating} 
\usepackage{comment}
\hypersetup{citecolor=red,colorlinks=true,urlcolor=blue}

\newcommand{\IOPPAS}{Institute of Physics PAS, Aleja Lotnik\'ow 32/46, 02-668 Warszawa, Poland}
\newcommand{\FUW}{Faculty of Physics, University of Warsaw, Pasteura 5, 02-093 Warszawa, Poland}
\newcommand{\baqis}{Beijing Academy of Quantum Information Sciences, Beijing 100193, China}
\newcommand{\TU}{State Key Laboratory of Low Dimensional Quantum Physics, Department of Physics, Tsinghua University, Beijing 100084, China}
\newcommand{\FSCQI}{Frontier Science Center for Quantum Information, Beijing, China}

\newcommand{\HNL}{Hefei National Laboratory, Hefei, Anhui 230088, China}

\begin{document}

\title{
Multisetting protocol for Bell correlated states detection with spin-$f$ systems
}

\author{Arkadiusz Kobus}
\affiliation{\IOPPAS}
\affiliation{\FUW}

\author{Xinwei Li}
\affiliation{\baqis}

\author{Mariusz Gajda}
\affiliation{\IOPPAS}

\author{Li You}
\affiliation{\baqis}
\affiliation{\TU}
\affiliation{\FSCQI}
\affiliation{\HNL}

\author{Emilia Witkowska}
\affiliation{\IOPPAS}
\email{ewitk@ifpan.edu.pl}

\begin{abstract}
We propose a multisetting protocol for the detection of two-body Bell correlations, and apply it to spin-nematic squeezed states realized in $f$ pairs of SU(2) subsystems within spin-$f$ atomic Bose–Einstein condensates. 
Experimental data for $f=1$, together with numerical simulations using the truncated Wigner method for $f=1,\,2,\,3$, demonstrate the protocol's effectiveness. We derive Bell inequalities tailored to collective observables of these subsystems and show how tuning interaction parameters controls the number of independent settings. Our findings extend multisetting Bell tests to high-spin ultracold ensembles, enabling scalable, flexible measurement configurations. This establishes a pathway to quantum metrology and information processing in tunable many-body platforms.
\end{abstract}

\date{\today}

\maketitle

\section{Introduction}
The correlations between measurement outcomes obtained in space separation are fundamentally bounded when described within classical frameworks~\cite{PhysicsPhysiqueFizika.1.195}.
These bounds are expressed as Bell inequalities~\cite{RevModPhys.86.419}, originally formulated for scenarios with two observers (parties), each selecting between two measurement settings. This formulation has provided the foundation for several key experimental tests~\cite{PhysRevLett.23.880,PhysRevLett.65.1838,doi:10.1126/science.1247715}.

Bell-correlated states, those capable of violating Bell inequalities, have found many practical applications in quantum information science~\cite{RevModPhys.86.419}.
Extending each observer's choice to more than two measurement settings imposes distinct constraints,
enabling more efficient protocols for enhanced violations~\cite{PhysRevLett.93.200401, PhysRevA.74.062109, Barasinski2021experimentally, PhysRevA.78.052103}, improving robustness against noise and experimental imperfections~\cite{PhysRevA.80.052116, Pironio2010, PhysRevA.110.052432}, and enriching range of applications in quantum information tasks
~\cite{PhysRevA.83.062123, PhysRevLett.83.3077, PhysRevA.64.030301, PhysRevLett.82.1052, Solmeyer_2018, PhysRevLett.88.137902}.
Existing multisetting protocols, however, are restricted to few parties and rely on the GHZ family of states, which restricts scalability and measurement flexibility~\cite{RevModPhys.86.419, ScaraniBook}.

Bell inequalities for multipartite entangled states that is characterized by by one- and two-body correlations in collective observables, have been theoretically defined and experimentally verified using bimodal squeezed Bose-Einstein condensates (BECs),
demonstrating scalability with the system size~\cite{doi:10.1126/science.1247715, PhysRevA.100.022121, doi:10.1126/science.aad8665, Schine2022}. 
Despite advances with squeezed BECs, the implementation of multisetting scenarios, in particular, those involving independent measurements, remains underaddressed.

In this work, we integrate multisetting Bell scenarios with multipartite squeezed states to benefit from their complementary advantages, namely scalability and flexibility in measurement configurations.
To this end, we study spin-$f$ BECs~\cite{KAWAGUCHI2012253,RevModPhys.85.1191} with $f\ge 1$.
The spin degree of freedom allows us to introduce $f$ pairs of SU(2) subsystems. 
Multipartite entangled states, in terms of squeezing, are 
generated via spin-mixing dynamics in each subsystem, independently. 
We analytically derive a subclass of permutationally invariant, device-independent Bell inequalities for these squeezed states and for an arbitrary number of settings~\cite{PRXQuantum.2.030329, PhysRevLett.119.170403}, defined by generators of subalgebras associated with the SU(2) subsystems, and each yielding two measurement results.
We demonstrate improvement of the protocol’s scalability when increasing numbers of atoms and measurement settings, and show how imposing specific conditions on interaction strengths enables control over the number of independent settings and flexibility in measurement configurations.

A key experimental challenge emerges from simultaneous measurements of distinct observables across different subsystems constituting settings, which was successfully addressed in recent experiments through a sequence of microwave
pulses~\cite{PhysRevLett.123.063603,cao2023jointestimationtwophasespin}.
Utilizing the experimental data for spin-1 BEC~\cite{cao2023jointestimationtwophasespin}, 
along with numerical simulations employing the truncated Wigner method (TWM) for $f=1,\,2,\,3$ 
we demonstrate the feasibility of multisetting Bell protocol, and its scalability and flexibility in the number and choice of measurement settings for experimentally relevant parameters.


\section{Squeezing within spin-$f$ BECs}

The spin-$f$ BEC is a system characterized by a $2f+1$ multicomponent order parameter, each corresponding to magnetic sublevels indexed by the magnetic quantum number $m=-f,\dots,f$ with $f$ being the integer spin of the atom.
The system is described by considering binary s-wave interactions that conserve the total spin $\mathcal{F}$ for a pair of atoms during their collisions~\cite{KAWAGUCHI2012253}.
The reduction of the spatial degrees of freedom, by using the single mode approximation (SMA) \cite{PhysRevLett.81.5257, PhysRevA.102.023324}, where all atoms share the same spatial wave function, leads to the Hamiltonian
\begin{equation}
\label{eq:higher spin}
        \hat{H}_{\rm int}=\left[c_1 \hat{F}^2+c_2\hat{A}_{0,0}^\dagger \hat{A}_{0,0}+c_3\sum_{\mathcal{M}=-2}^2 \hat{A}_{2,\mathcal{M}}^\dagger\hat{A}_{2,\mathcal{M}}\right],
\end{equation}
for $f=1,2,3$, where
\begin{equation}
        \hat{A}_{\mathcal{F},\mathcal{M}}=\sum_{m=-f}^f\langle f,m,f,\mathcal{M}-m|\mathcal{F},\mathcal{M}\rangle \hat{a}_{m}\hat{a}_{\mathcal{M}-m},
\end{equation}
and $\langle \mathcal{F}, \mathcal{M}|f,m;f,m'\rangle$ are the Clebsch–Gordan coefficients while $\hat{a}_m$ are annihilation operators of an atom at the magnetic level $m$. The interaction coefficients $c_{1/2/3}$ are associated with the scattering lengths of colliding pairs of atoms with total spin $\mathcal{F}$~\cite{KAWAGUCHI2012253, RevModPhys.85.1191}. 
In Eq.\eqref{eq:higher spin}, the spin-$f$ operator is given by $\hat{F}^2 = \hat{F}_x^2+\hat{F}_y^2 + \hat{F}_z^2$ and the interaction coefficients satisfy $c_1\ne0 $ and $c_{2/3}=0$ for $f=1$, $c_{1/2}\ne0$ and $c_3=0$ for $f=2$, and $c_{1/2/3}\ne 0$ for $f=3$. 
When an external magnetic field is included, the total Hamiltonian becomes
\begin{equation}
\label{eq:ham}
    \hat{H} = \hat{H}_{\rm int} - q \sum_{m=-f}^f m^2 \hat{N}_m,
\end{equation}
where $q$ denotes the quadratic Zeeman energy coefficient~\cite{Ho1998,Machida,PhysRevA.88.033629}, $\hat{N}_m$ is the atom-number operator for the Zeeman sublevel $m$, and constant terms have been omitted
\footnote{The contribution of linear Zeeman energy is irrelevant as it is proportional to the magnetization $M=N_1-N{-1}$ which is a constant of motion~\cite{PhysRevA.90.043609}. The quadratic Zeeman energy is of the main importance in the lowest order approximation for realistic systems}.

\begin{figure}[]
    \centering
    \includegraphics[width=0.45\textwidth]{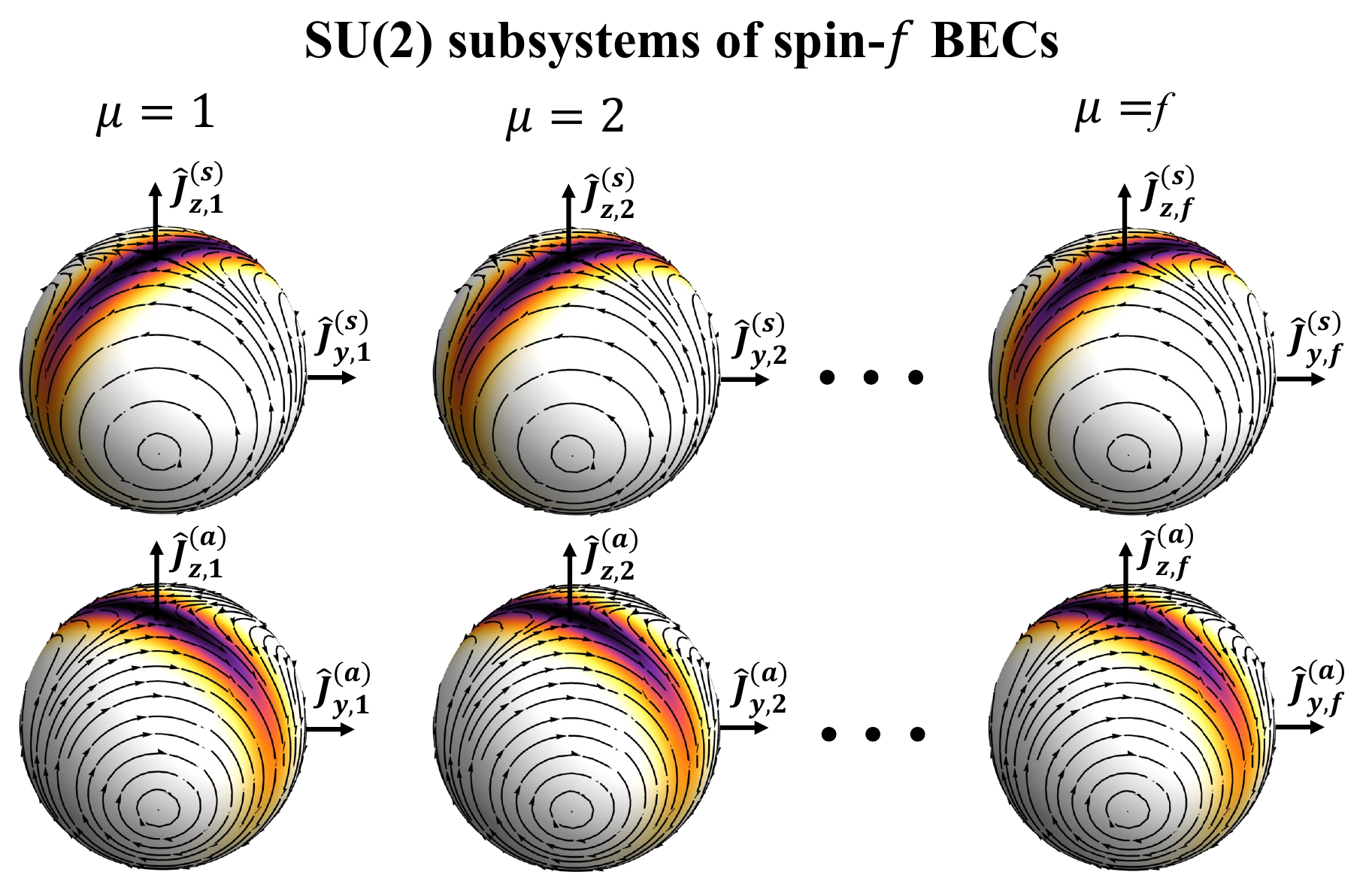}
   \caption{The generalized Bloch spheres in SU(2) subsystems are spanned by the triplet of operators $\{ \hat{J}^{(\sigma)}_{x,\mu}, \hat{J}^{(\sigma)}_{y,\mu}, \hat{J}^{(\sigma)}_{z,\mu} \}$ which are defined in Eq.(\ref{eq:spin operatorsx})-(\ref{eq:spin operatorsz}). There are $f$ ($\mu = 1,\, 2,\cdots, f$) pairs of subsystems, each corresponding to a symmetric (\(\sigma = s\)) and an antisymmetric (\(\sigma = a\)) component. On each sphere, ‌constant-energy contours‌ under the mean-field approximation are plotted as black curves, with arrows indicating the ‌direction of time evolution. The Husimi function is represented by color, illustrating the squeezed states generated simultaneously in each subsystem.}
    \label{fig:fig1}
\end{figure}

The spin degrees of freedom allow us to define a collection of $f$ pairs of subsystems, as illustrated in Fig.~\ref{fig:fig1}. 
We introduce the symmetric and anti-symmetric annihilation operators $\hat{g}^{(s)}_\mu=(\hat{a}_\mu+\hat{a}_{-\mu})/\sqrt{2}$ and $\hat{g}^{(a)}_\mu=(\hat{a}_\mu-\hat{a}_{-\mu})/\sqrt{2}$, as well as the corresponding pseudo-spin operators:
\begin{align}
\label{eq:spin operatorsx}
&\hat{J}^{(\sigma)}_{x,\mu}
=\frac{1}{2}(\hat{g}^{(\sigma)}_\mu{}^\dagger \hat{a}_0
+\hat{a}^\dagger_0 \hat{g}^{(\sigma)}_\mu), \\
&\hat{J}^{(\sigma)}_{y,\mu}=
\frac{i}{2}(\hat{g}^{(\sigma)}_\mu{}^\dagger\hat{a}_0-\hat{a}^\dagger_0 \hat{g}^{(\sigma)}_\mu), \\
&\hat{J}^{(\sigma)}_{z,\mu}=\frac{1}{2}(\hat{a}^\dagger_0\hat{a}_0-\hat{g}^{(\sigma)}_\mu{}^\dagger \hat{g}^{(\sigma)}_\mu),
\label{eq:spin operatorsz}
\end{align}
where the indices $\sigma=s$ and $\sigma=a$ refer to the symmetric and anti-symmetric subspaces, respectively, within the given absolute value of the magnetic number sector $\mu = |m|$~\cite{PhysRevA.46.R6797,PhysRevA.88.033629,Yukawa_2016}.
For $f=1$, there is one pair of symmetric and anti-symmetric spin operators ($\mu = 1$). For $f=2$, there are two pairs ($\mu =1, 2$), and for $f=3$, there are three pairs ($\mu =1, 2, 3$). 
The operators in Eq.(\ref{eq:spin operatorsx})-(\ref{eq:spin operatorsz}) form SU(2) subalgebras that satisfy the canonical commutation relations, e.g., $[\hat{J}^{(\sigma)}_{x,\mu}, \hat{J}^{(\sigma)}_{y,\mu}]= i \hat{J}^{(\sigma)}_{z,\mu}$ for any $\sigma$.

Spin-nematic squeezing is generated independently in each subsystem, starting from the initial coherent state 
$ |\psi_0\rangle = \frac{\hat{a}_0^\dagger{}^N}{\sqrt{N!}} |0\rangle $, 
where $|0\rangle$ denotes the vacuum state.
The first-order binary interactions between atoms in the zero magnetization channel ($\mathcal{M}=0$) alter the population of magnetic sublevels by generating correlated atom pairs in the $m$ and $-m$ states from colliding atoms in the $m=0$ magnetic level, and vice versa~\cite{Hamley2012,Liu2022}. These interactions drive spin-mixing dynamics, leading to the formation of two-body correlations in the system
~\footnote{Higher-order binary interaction processes occur on longer timescales, provided that the total spin $\mathcal{F}$ of the colliding atomic pair is conserved during collisions. These processes result in slight variations in the rates and levels of squeezing generation across different subsystems when $f>1$, as discussed in Section~\ref{appsec:Spin-2 and spin-3 Becs}.}.

The level of squeezing is characterized by the squeezing parameter
\begin{equation}
	\xi_{\mu, \sigma}^2 = \frac{N(\Delta \hat{J}^{(\sigma)}_{\mathrm{min},\mu})^2}{|\langle \hat{J}^{(\sigma)}_{z,\mu} \rangle|^2},
\end{equation}
where $(\Delta \hat{J}^{(\sigma)}_{\mathrm{min},\mu})^2$ represents the minimal variance in the plane orthogonal to the direction of the mean collective spin $\langle \vb*{J}^{(\sigma)}_\mu\rangle$. Here, $\vb*{J}^{(\sigma)}_\mu = (\hat{J}^{(\sigma)}_{x,\mu}, \hat{J}^{(\sigma)}_{y,\mu}, \hat{J}^{(\sigma)}_{z,\mu})$ for each $\mu =1,2,\dots, f$ and $\sigma=s,a$. 
Illustrative examples of the time evolution of the squeezing parameters under the Hamiltonian in Eq.~(\ref{eq:ham}) are provided in Appendix~\ref{appsec:Spin-2 and spin-3 Becs} for the cases $f=2$ and $f=3$.

It is worth noting that nearly all atoms remain in the $m=0$ state on short timescales comparable to the optimal squeezing time. As a result, the mean spin remains directed along $\hat{J}^{(\sigma)}_{z,\mu}$ on this short timescale, and the mean value of the spin length is uniform across all subspaces labeled by $\sigma$ and $\mu$, such that $\langle \hat{J}^{(\sigma)}_{z,\mu}\rangle=\langle \hat{J}^{(\sigma')}_{z,\mu'}\rangle$. Therefore, it is denoted by $\hat{J}_{z,\mu}$ from now on.

\section{Bell inequality for spin-nematic squeezed states od spin-$f$ systems}
\label{sec:Bellinequality}

We employ a data-driven approach to derive the corresponding Bell inequality. 
A pedagogical discussion of the derivation, together with an explicit presentation of the intermediate steps and the final expressions, is provided in Appendix~\ref{app:Bellscenario}. Here we present only the key results.

The Bell inequality, compatible with local-variable theory for any input data $\vec{M}$ (containing one-body correlators) and $\tilde{C}$ (covariance of two-body correlations), as well as for any positive semi-definite matrix $A$ and vector $\vec{h}$, can be defined using the data-driven method~\cite{PRXQuantum.2.030329,PhysRevA.111.023312},
\begin{equation}
    L(A, \vec{h})=E_{\mathrm{max}}(A, \vec{h}) + \vec{h} \cdot \vec{M} + \mathrm{Tr}[A \tilde{C}]\ge 0,
    \label{eq:generalBellinequality}
\end{equation}
where the classical bound is set by ${E_{\mathrm{max}}(A, \vec{h})=N\max_{\vec{r}\in \{\pm 1/2\}}[\vec{r}^T A\vec{r} - \vec{h}\cdot\vec{r}]}$ for two measurement outcomes. 
Let us start with the choice of convenient measurement settings. Our analysis of various observables reveals that the most relevant is the collective operator,
\begin{align}
    \hat{J}_{\alpha,\mu} &=  
    \hat{J}^{(s)}_{\mathrm{min},\mu}
    \sin\theta_{\alpha,\mu}\cos\varphi_{\alpha,\mu}
     + \hat{J}^{(a)}_{\mathrm{min},\mu}
     \sin\theta_{\alpha,\mu}\sin\varphi_{\alpha,\mu} \nonumber \\
    &+\hat{J}_{z,\mu}
    \cos\theta_{\alpha,\mu},
    \label{eq:settingschosen}
\end{align}
where each measurement setup $(\alpha,\mu)$ corresponds to selecting a subspace $\mu$ and a direction $\alpha$ in the subspace spanned by $(\hat{J}^{(s)}_{\mathrm{min},\mu},\hat{J}^{(a)}_{\mathrm{min},\mu},\hat{J}_{z,\mu})$. 
The measurement outcomes are binary, given by $r_{\alpha,\mu}^{(j)}=\pm 1/2$. The number of directions $\alpha$ is associated with the number of settings chosen by the observer within the given subsystem $\mu$, with $\alpha = 1,2,\dots, k_\mu$. The total number of possible measurements is $k=\sum_{\mu=1}^fk_\mu$.
Therefore, we introduce the input vector $\vec{M}$ and the input matrix $\tilde{C}$ whose elements are 
\begin{align}
    \frac{M_{\alpha,{\mu}}}{N} &= 
    v_\mu \cos\theta_{\alpha,\mu},
\end{align}
with $v_\mu=\langle \hat{J}_{z,\mu} \rangle/N$, and
\begin{align}
\label{eqM:C matrix}
    &\frac{\tilde{C}^{\alpha,\mu}_{\alpha',\mu'} }{N}
    =\delta_{\mu,\mu'}\left[- \frac{1}{4}
    \cos\theta_{\alpha,\mu} 
    \cos\theta_{\alpha',\mu'}\right.  \nonumber\\
    &+
    \left(
    v_\mu ^2 \xi^2 _{\mu, s} - \frac{1}{4} \right)\sin\theta_{\alpha,\mu} \sin\theta_{\alpha',\mu'} \cos\varphi_{\alpha,\mu}
    \cos\varphi_{\alpha',\mu'} \nonumber \\
    &+ 
    \left.\left(v_\mu^2 \xi^2 _{\mu, a} - \frac{1}{4} \right)
    \sin\theta_{\alpha,\mu} \sin\theta_{\alpha',\mu'} 
    \sin\varphi_{\alpha,\mu}  \sin\varphi_{\alpha',\mu'}\right],
\end{align} 
see Appendix~\ref{app:Bellscenario} for further details.
Note that the matrix $\tilde{C}$ given by Eq.~(\ref{eqM:C matrix}) has a block-diagonal structure with $f$ blocks indexed by $\mu$. Each block is a $k_\mu\times k_\mu$ matrix with elements labeled by $\alpha$ and $\alpha'$.

The minimization of $L(A, \vec{h})$ for the squeezed states, with $\xi^2_{\mu, \sigma}<1$, is performed to identify the optimal matrix $A$ and vector $\vec{h}$, giving $A_{\alpha,\mu;\alpha',\mu'}=4\delta_{\mu,\mu'}w_\mu$ and $h_{\alpha,\mu}=4w_\mu(2\alpha-1-k_\mu)$ for all $\alpha,\mu,\alpha',\mu'$, where $w_\mu\ge0$ are weights between blocks. 
We have $A_{\alpha,\mu;\alpha',\mu'}=0$ and $h_{\alpha,\mu}=0$ whenever $\xi^2_{\mu, \sigma}\ge 1$.
The classical bound is given by $E_{\mathrm{max}}=N\sum_{\mu=1}^fw_\mu k_\mu^2$.
The resulting Bell inequality takes the form of a weighted sum:
\begin{align}
   E_{\mathrm{max}}^{-1}
   \sum_{\mu=1}^fw_\mu k_\mu^2\mathcal{L}(k_\mu,\vec{\theta}_\mu,\vec{\varphi}_\mu)\ge0,
    \label{eqM:generalBE}
\end{align}
with
\begin{align}
   \mathcal{L}(k_\mu,\vec{\theta}_\mu,\vec{\varphi}_\mu)
   &=
   1 + k_\mu^{-2}v_\mu \sum_{\alpha=1}^{k_\mu}h_{\alpha,\mu}\cos\theta_{\alpha,\mu} 
   - k_\mu^{-2} g_{\vec{\theta}_\mu}^{(c)}\nonumber \\
    & - k_\mu^{-2}(1-4v_\mu^2\xi_{\mu, s}^2)
    g^{(sc)}_{\vec{\theta}_\mu,\vec{\varphi}_\mu} \nonumber \\
    & - k_\mu^{-2}(1-4v_\mu^2\xi_{\mu, a}^2)
    g^{(ss)}_{\vec{\theta}_\mu,\vec{\varphi}_\mu}.
\end{align}
The functions $g$ are defined as:
\begin{align*}
    g^{(sc)}_{\vec{\theta}_\mu,\vec{\varphi}_\mu} 
        &= \left(\sum_{\alpha=1}^{k_\mu} \sin\theta_{\alpha,\mu} \cos\varphi_{\alpha,\mu} \right)^2, \\
    g^{(ss)}_{\vec{\theta}_\mu,\vec{\varphi}_\mu} 
        &= \left(\sum_{\alpha=1}^{k_\mu} \sin\theta_{\alpha,\mu} \sin\varphi_{\alpha,\mu}\right)^2, \\
    g_{\vec{\theta}_\mu}^{(c)} 
        &= \left(\sum_{\alpha=1}^{k_\mu} \cos\theta_{\alpha,{\mu}} \right)^2.
\end{align*}
\begin{figure}[]
    \centering
    \includegraphics[width=\linewidth]{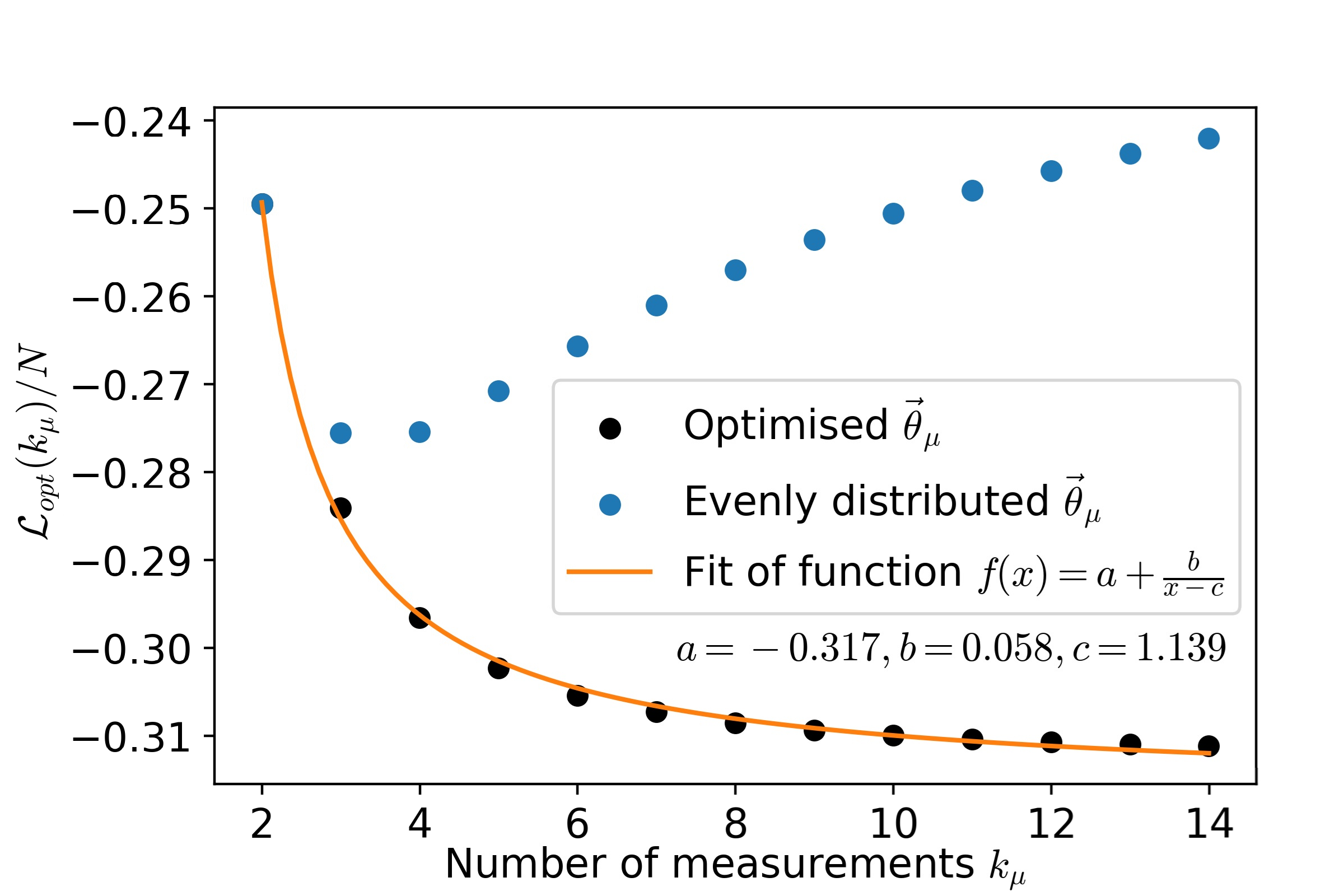}
    \caption{
    The Bell correlator in the given subsystem $\mu$, normalized by the total number of atoms, $\mathcal{L}_\text{opt}(k_\mu)/N$, versus the number of measurement settings $k_\mu$ when angles $\vec{\theta}$ are optimal given by Eqs. (\ref{eqM:thetasigma}) and (\ref{eqM:thetasigma2}) (marked by black point) and evenly distributed in the region $[\gamma,\pi-\gamma]$ for $\gamma\approx0.22\pi$ (blue points). The orange solid line represents the function with the fitted parameters listed in the legend‌.}
    \label{figM:fig6}
\end{figure}

Finally, we minimize $\mathcal{L}(k_\mu,\vec{\theta}_\mu,\vec{\varphi}_\mu)$ over the angles $\vec{\theta}_{\mu}$ and $\vec{\varphi}_{\mu}$ within each block $\mu$.
The minimization with respect to $\varphi_\alpha$ yields two sets of solutions:
$\cos\varphi_\alpha=1$ if $\xi_{\mu, s}^2<\xi_{\mu, a}^2$,
and
$\cos\varphi_\alpha=0$ if $\xi_{\mu, a}^2<\xi_{\mu, s}^2$, with the smaller value denoted by $\xi^2_\mu$ henceforth.
This suppresses the contribution from the larger squeezing parameter.
The minimization over the angles $\vec{\theta}_\mu$ gives the following relations:
\begin{align}
\label{eqM:thetasigma}
    \theta_{\alpha,\mu } &= 
    \pi - \theta_{k_\mu+1-\alpha,\mu},\\
    \label{eqM:thetasigma2}
    \frac{k_\mu^2 v_\mu h_{\alpha,\mu}}{2(1-4 v_\mu^2 \xi_\mu^2)} 
    \tan\theta_{\alpha,\mu} &= - \sum_{\alpha' = 1}^{k_\mu} \sin\theta_{\alpha',\mu}.
\end{align}
The angle-optimized Bell correlator is given by
\begin{equation}
    \frac{L^{(k)}_\text{opt}}{E_{\mathrm{max}}}=\frac{\sum_{\mu=1}^fw_\mu k_\mu^2 \mathcal{L}_\text{opt}(k_\mu)}{\sum_{\mu=1}^fw_\mu k_\mu^2},
    \label{eqM:BelloverEmaxgeneral}
\end{equation}
where the values of $\mathcal{L}_\text{opt}(k_\mu)$ are shown in Fig.~\ref{figM:fig6} for parameters in the thermodynamic limit, when $v_\mu=1/2$ and $\xi_\mu^2 \to 0$. One can see that for this case, the value of $\mathcal{L}_\text{opt}(k_\mu)$ decreases with $k_\mu$, saturating approximately at $-0.317N$. This demonstrates enhancement due to the increased number of settings and scalability with the number of atoms $N$. 

Optimization over weighting parameters $w_\mu$ selectively enhances contributions from subsystems $\mu$ where the optimized $k_\mu^2 \mathcal{L}_\text{opt}(k_\mu)$ exhibits substantial negative values, while suppressing contributions from subsystems where $k_\mu^2 \mathcal{L}_\text{opt}(k_\mu)>0$.
If $k_\mu^2 \mathcal{L}_\text{opt}(k_\mu)$ is negative and exhibits weak $\mu$-dependence, employing a uniform weighting strategy proves advantageous, as it increases the contributions from all subsystems, effectively utilizing a larger total number of settings $k$. 

Note that the multisetting two-body Bell correlator presented in~\cite{PhysRevLett.119.170403} represents a special case of the left-hand side of Eq.~(\ref{eqM:generalBE}) for a given subsystem $\mu$, under the assumption that $\xi_{\mu, s}^2=\xi_{\mu, a}^2$.

\section{Results for $f=1,2,3$ BECs}
\label{sec:results}

The application of the Bell inequality to spin-$f$ BECs, including a detailed account of the intermediate steps and the resulting expressions, is presented in Appendices~\ref{sec:Spin-1 BECs.-} and \ref{appsec:Spin-2 and spin-3 Becs}. They serve to elucidate the methodology and to provide clarity for readers less acquainted with the technical aspects of the Bell inequality applications. Below, we focus solely on the key results.

We begin by exploring the above theory for $f=1$ using experimental data obtained by measuring squeezing in the symmetric and antisymmetric subspaces simultaneously~\cite{cao2023jointestimationtwophasespin}. 
The corresponding Bell inequality takes the form
\begin{equation}
\label{eqM:thetavarphiN}
\frac{L^{(3)}_{{\rm opt}}}{E_{\rm max} } =1-\frac{16}{9}v\cos\theta_{3\mathrm{opt}}-\frac{1-4v^2\xi_1^2}{9}(1+2\sin\theta_{3\mathrm{opt}})^2, 
\end{equation}
where the optimal angle $\theta_{3\mathrm{opt}}$ satisfies 
$\frac{4v}{1 - 4v^2\xi_1^2} \tan\theta_{3\mathrm{opt}} = 1 + 2\sin\theta_{3\mathrm{opt}}$.
In the thermodynamic limit, $\theta_{3\mathrm{opt}}=\arctan(\sqrt{5/3}) \approx 0.91$, yielding 
\begin{equation}
L^{(3)}_{\rm opt}/E_{\rm max} \bigg|_{N\to \infty} \approx -0.285,
\end{equation}
which represents a $12\%$ enhancement in Bell violation compared to the two-measurement scenario with $L^{(2)}_{\rm opt}/E_{\rm max}\approx -0.25$~\cite{PhysRevA.111.023312}.

\begin{figure}[]
    \centering
    \includegraphics[width=0.95\linewidth]{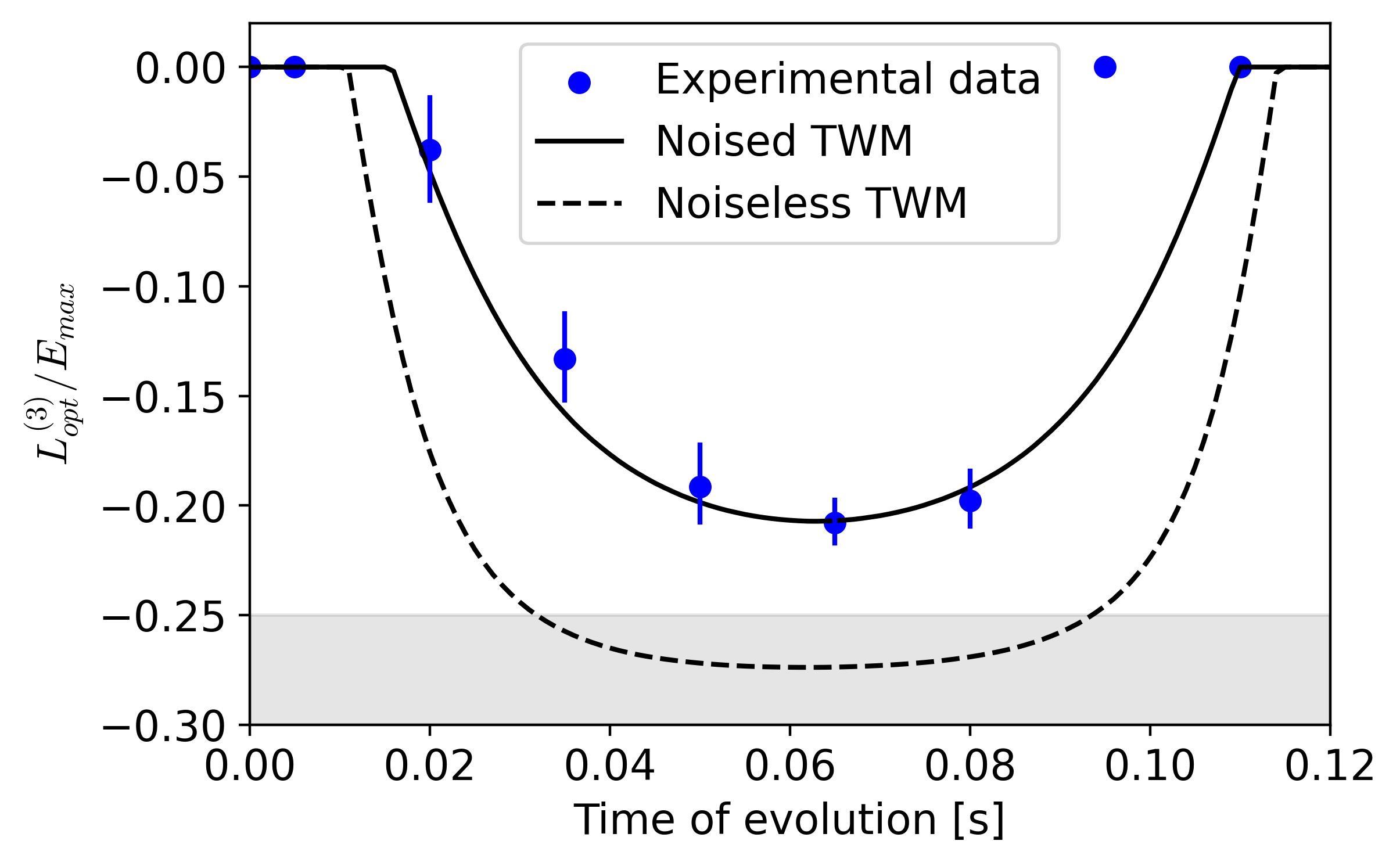}
    \caption{
    The evolution of the Bell correlator with three measurement settings $L^{(3)}_{{\rm opt}}/E_{\rm max}$ given by Eq.~(\ref{eqM:thetavarphiN}). The blue points mark the results obtained using the experimental data in Ref.~\cite{cao2023jointestimationtwophasespin}, where squeezing parameters $\xi^2_{1, s} $ and $\xi^2_{1, a}$ were measured simultaneously for $N=26500$ and $q=|c_1|$. 
    Numerical simulations using the TWM are shown with (solid line) and without (dashed line) additional noise from microwave measurement pulses, respectively. The shaded region indicates enhancement with respect to two measurement settings.}
    \label{fig:fig7}
\end{figure}

In Fig.~\ref{fig:fig7}, we show the time evolution of the Bell correlator in Eq.~(\ref{eqM:thetavarphiN}), comparing experimental data from spin-1 rubidium Bose-Einstein condensates~\cite{cao2023jointestimationtwophasespin} with TWM simulations. 
Experimental points (blue dots) represent simultaneous measurements of squeezing parameters $\xi^2_{1,s}$ and $\xi^2_{1,a}$, while numerical results with (solid line)
and without (dashed line) additional noise from the implementation of microwave measurement pulses are shown~\footnote{For the non-ideal case of the TWM, we use the following set of parameters: $\delta q=3.8\cdot 10^{-3}$, $\gamma=0.069 \text{ Hz}$, $\Delta=24$, $\eta=0.3$, $\delta\eta=4.7\cdot10^{-2}$. See Supplementary Materials of Ref.~\cite{Liu2022, cao2023jointestimationtwophasespin} for the meanings of these notations}. 
The TWM maps the field operators $\hat{a}_m$ and $\hat{a}_m^\dagger$ to complex stochastic variables $\alpha_m$ and $\alpha_m^*$, transforming the system's dynamics into a set of stochastic differential equations~\cite{sinatra2002truncated, szigeti2017pumped}, which enables simulation for macroscopic numbers of atoms including technical noise. The corresponding numerical codes are available in~\cite{repository}. The results demonstrate the implementation of three-measurement settings, although experimental imperfections limit the values of the Bell correlator, and enhancement over the two-measurement setting is not observed (shaded area).

\begin{figure}[] 
    \centering
    \includegraphics[width=0.95\linewidth]{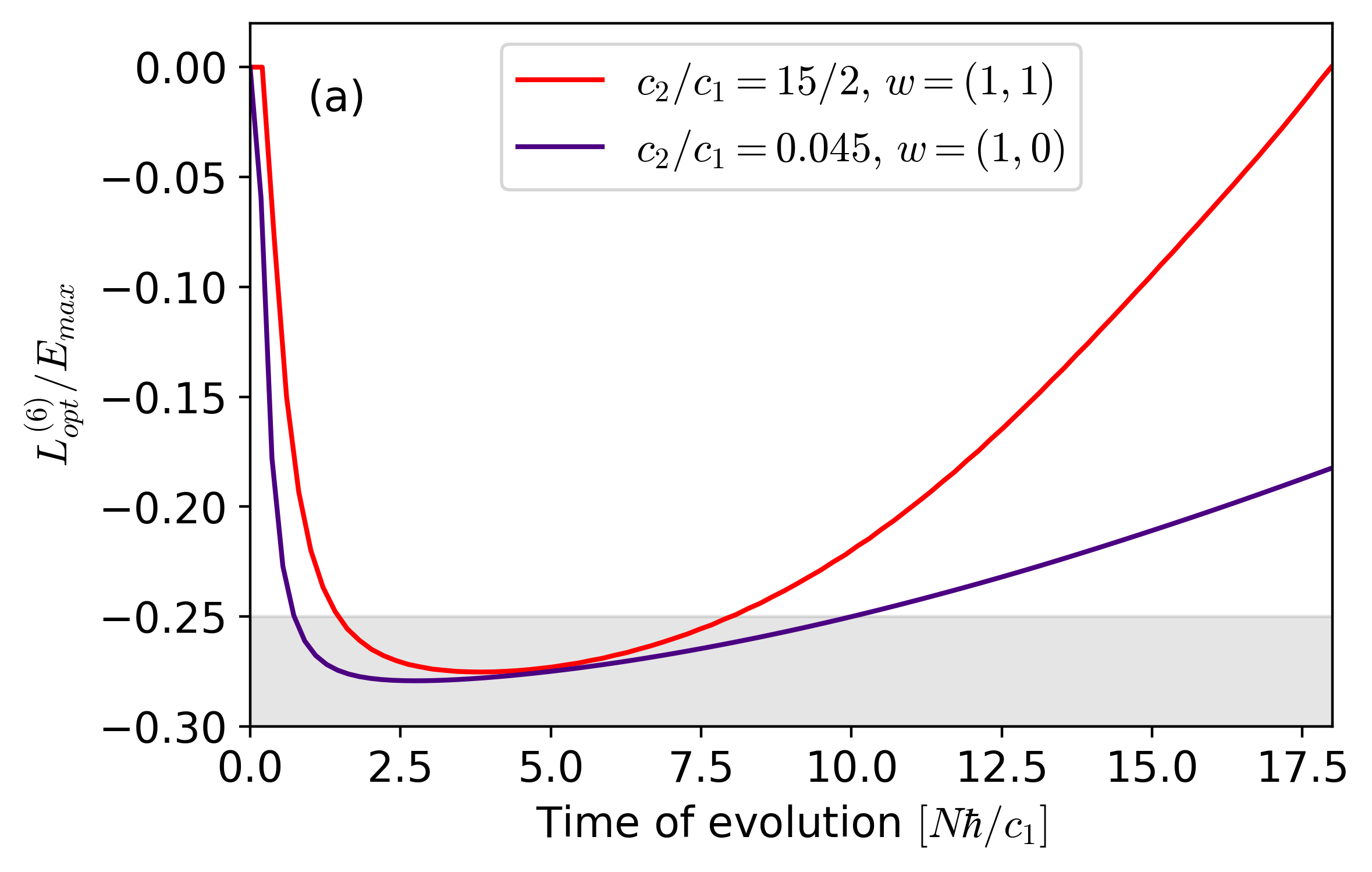}
     \includegraphics[width=0.95\linewidth]{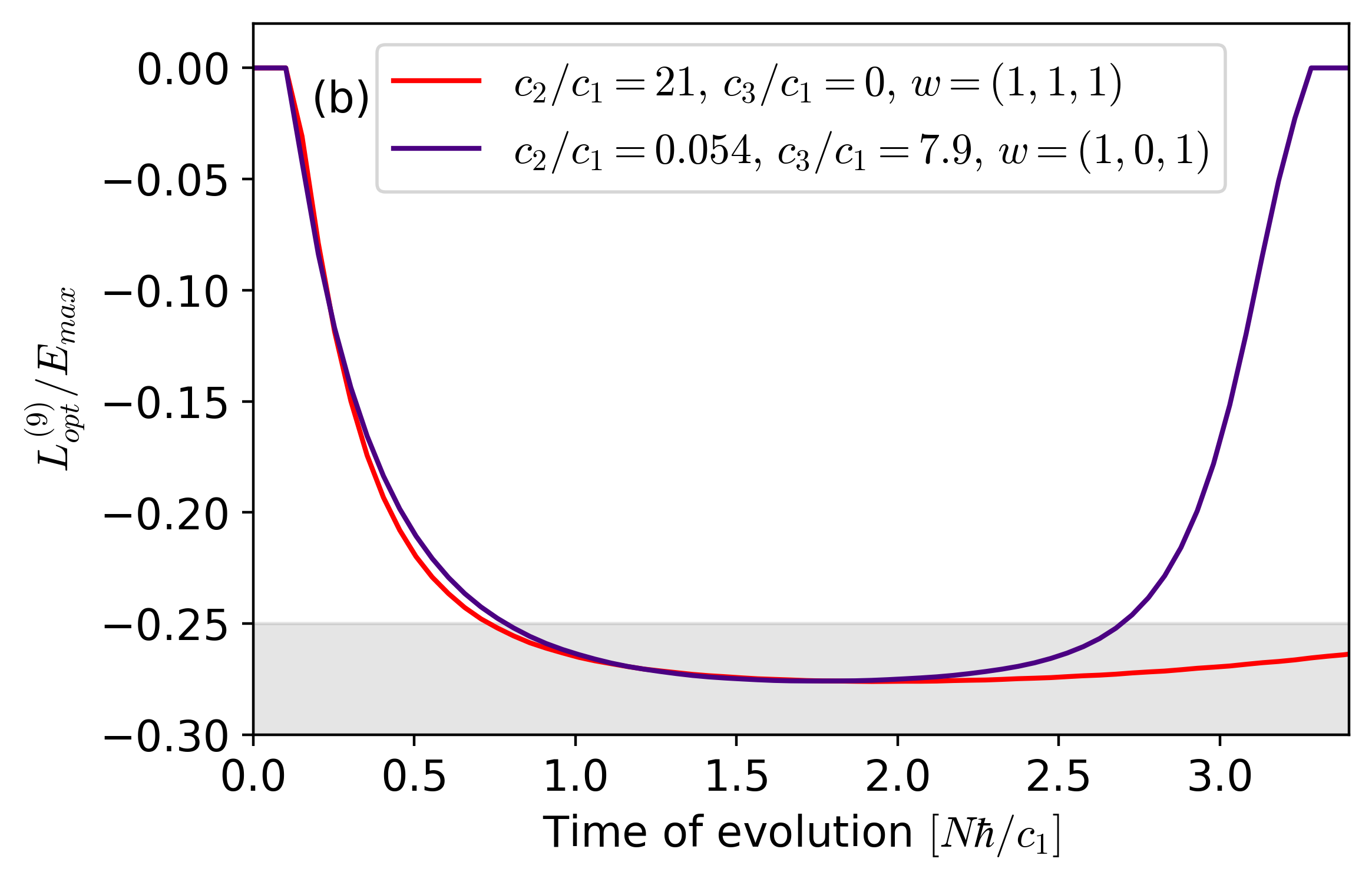}
    \caption{The Bell correlator~(\ref{eqM:BelloverEmaxgeneral}) given by the evolution using the noiseless TWM for $N=26400$, $q=c_1$ and $f=2$ (a) and $f=3$ (b). The blue curves correspond to the interaction coefficients of Rb and Cr atoms with their natural scattering lengths~\cite{KAWAGUCHI2012253, Widera_2006, Chomaz_2023}. The red curves correspond to the adjusted scattering length required to achieve equal initial squeezing rates across all subsystems. The weights of each block $w_\mu$ are listed in legend.}
    \label{fig:fig8}
\end{figure}

The Bell correlator for larger spin systems is presented in Figs.~\ref{fig:fig8}. 
The scattering lengths of Na and Rb ($f=2$) are
$|c_2/c_1|\approx 1.57$ 
and $|c_2/c_1|\approx 0.045$,
respectively~\footnote{The scattering lengths are  $a_0\approx 34.9 a_B$, $a_2\approx 45.8 a_B$, and $a_4\approx 62.51 a_B$ for sodium-23 atoms~\cite{KAWAGUCHI2012253}; and $a_0\approx 87.93 a_B$, $a_2\approx 91.28 a_B$, and $a_4\approx 98.98 a_B$ for rubidium-87 atoms~\cite{Widera_2006}, both with spin $f=2$. Here, $a_B$ is the Bohr radius.}. 
These values result in significant squeezing only in the $\mu=1$ subsystem, enabling an effective three-setting Bell protocol, as shown by the violet line in Fig.~\ref{fig:fig8}(a).
For Cr atoms ($f=3$), we have $|c_2/c_1| \approx 0.054$ and $|c_3/c_1| \approx 7.9$
~\footnote{These values correspond to scattering lengths $a_0\approx 13.5 a_B$, $a_2\approx -7 a_B$, $a_4\approx 56 a_B$, and $a_6\approx 102.5 a_B$ taken from~\cite{Chomaz_2023}}, 
indicating non-negligible squeezing in both the $\mu=1$ and $\mu=3$ subsystems. This allows for an effective six-setting Bell protocol, as shown in Fig.~\ref{fig:fig8}(b).
Careful tuning of scattering lengths is necessary to achieve the maximum number of independent settings for a given $f$.
We derive analytical relationships between the parameters $c_1,\, c_2,\, c_3$ that produce equal rates of initial squeezing dynamics across all subsystems $\mu$, yielding  
$(0,c_2,0)$ and $(4c_2/30,c_2,0)$ for $f=2$;
and 
$(0,c_2,0)$, $(c_2/21,c_2,0)$,
$(c_2/35,c_2,12c_2/10)$, and $(8c_2/105,c_2,12c_2/10)$ for $f=3$, where the values in the brackets refer to $(c_1,c_2,c_3)$; see Appendix~\ref{appsec:Spin-2 and spin-3 Becs} for details. The corresponding Bell correlator for such parameters is shown by red lines in Figs.~\ref{fig:fig8}. These results underscore the necessity of fine-tuning coupling coefficients to balance contributions from distinct channels $\mathcal{M}$. They also indicate that our protocol accommodates flexibility in choosing the number of measurement settings.

\section{Conclusions}

In conclusion, we report a multisetting protocol for detecting Bell-correlated states using spin-$f$ BECs, and discuss its application to experimental data~\cite{Liu2022} and numerical results obtained with the TWM method using spin-nematic squeezing generated in the $f$ pairs of SU(2) subsystems. The latter can be used in multiparameter estimation tasks~\cite{Demkowicz_Dobrza_ski_2020,cao2023jointestimationtwophasespin,li2025multiparameterestimationarrayentangled}. 
The Bell correlations detected can be even larger than those estimated by Eq.~(\ref{eqM:BelloverEmaxgeneral}) when additional correlations between the $\sigma=s$ and $\sigma=a$ subsystems are present. The required processes are not inherently generated by the spin-$f$ Hamiltonian, and we provide an example in Appendix~\ref{app:mixing}. 

The multisetting Bell protocol can be realized in other ultracold atomic systems, including Yb and Sr platforms, which are currently being investigated for quantum information applications~\cite{PhysRevA.78.022301}.
The interactions necessary for squeezing generation can be engineered using Floquet techniques~\cite{doi:10.1126/science.abd9547}. Our protocol integrates the benefits of multisetting features and multipartite entanglement, which were previously addressed independently. We demonstrate how to incorporate both within a single protocol and highlight the potential scalability and flexibility through multipartite squeezed states.
\\
\section*{ACKNOWLEDGMENT}
We acknowledge discussions with R. Augusiak and M. Fadel. This work is supported by the Polish National Science Centre SHENG project DEC-2023/48/Q/ST2/00087, and the National Natural Science Foundation of China (NSFC) (Grants No. 92265205 and No. 12361131576).

\appendix

\section{Tailoring Bell inequality for squeezed spin-f BECs}
\label{app:Bellscenario}

\begin{figure}[]
	\centering
	\includegraphics[width=0.45\textwidth]{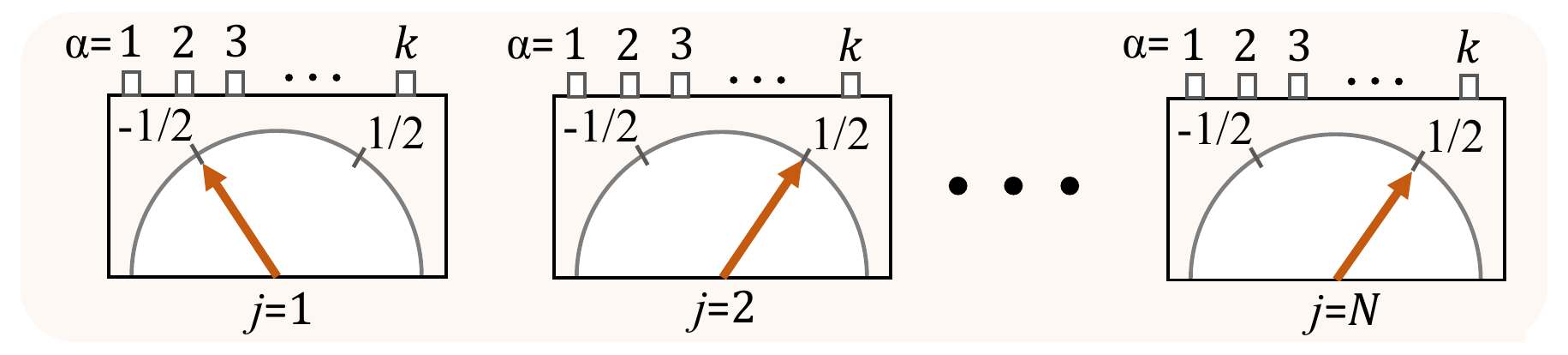}
	\caption{
		The entangled state $\hat{\rho}$ is shared by $N$ parties. For each party, one can choose between $k$ local measurement settings. For each party and setting chosen, there are two measurement results possible $\pm 1/2$. In the case of spinor BEC, each party corresponds to an individual spin-$f$ boson‌ that shares ‌a squeezed state‌ within its respective subsystem $\mu$. The measurement settings $(\alpha, \mu)$ are defined in the respective subsystems (\ref{eq:s01}), each leading to the two measurement outcomes.
	}
	\label{fig:fig5}
\end{figure}

In this work, we restrict our analysis to a subclass of Bell inequalities, known as two-body permutationally invariant Bell inequalities, which are constrained by symmetries and involve at most two-body correlation functions.
We consider $N$ parties in the Bell scenario. For each party $j=1,\cdots,N$, one can choose among $k$ local measurement settings labeled by $\alpha=1,2,\cdots,k$. 
The scenario considered here involves choosing a set of observables $\alpha=\{\alpha_j\}_{j=1}^N$ for each of the $N$ subsystems and recording the local measurement results $r=\{r_{j,\alpha}\}_{j=1}^N$, where each $r_{j,\alpha}=\pm 1/2$ as illustrated in Fig.~\ref{fig:fig5}.
In the local hidden variable (LV) theory, the probability distribution $P_N(r|\alpha)$ for the outcomes $r$ given the settings $\alpha$ can be written as
\begin{align}\label{eq: DistProbLocal}
	P_N^{(\rm LV)}(r|\alpha)=\int d\lambda\,q(\lambda)\prod_{j=1}^MP_\lambda^{(j)}(r_j|\alpha_j),
\end{align}
for all possible choices of $\alpha$.
If the measured probability distribution $P_N(r|\alpha)$ cannot be written in the form (\ref{eq: DistProbLocal}), the correlations present in the system are nonlocal. The locality here means that the probability distribution of the outcomes for any given subsystem $j$ depends only on the setting within the same subsystem $j$, leading to $P_\lambda^{(j)}(r_{j,\alpha}|\alpha_j)=\delta_{r_{j,\alpha},r_{j,\alpha}(\lambda)}$
~\footnote{In other words, the delta probability distribution holds if we consider a local realistic model where the local measurement result $r_{j,\alpha}$ is deterministically determined by the setting $\alpha_j$ and the variable $\lambda$ in each subsystem $j$ independently.}. 
Instead of working with a probability distribution, it is equivalent to consider the expectation values for the average measurement outcomes and the product of two measurement outcomes, $\langle r_{j,\alpha} \rangle$ and $\langle r_{j,\alpha}r_{j',\alpha}\rangle$.

We use a data-driven method to derive Bell inequalities tailored for systems with an arbitrary number of measurements.
In the method, one introduces the vector ${\vec{M}=(M_1, \cdots, M_{k})}$ and matrix $\tilde{C}$ whose elements are 
\begin{align}\label{eq:AverageSpin}
	M_{\alpha} &= \sum_j \langle r_{j, \alpha} \rangle \\ \label{eq:CorrelationsSpin}
	\tilde{C}_{\alpha\alpha'} &=\sum_{j,j'\neq j}\langle r_{j, \alpha}r_{j',\alpha'}\rangle
	- M_{\alpha}M_{\alpha'},
\end{align} 
with $\alpha,\alpha'=1,\cdots,k$. 
It can be shown that the following Bell inequality holds
\begin{equation}
	L(A, \vec{h})=N E_{\rm max}(A, \vec{h}) + \vec{h} \vec{M} + Tr[A \tilde{C}]\ge 0,
	\label{eq:generalBellinequality}
\end{equation}
where the classical bound is set by ${E_{\rm max}(A, \vec{h})={\rm max}_{\vec{r}\in \{\pm 1/2\}}[\vec{r}^T A\vec{r} - \vec{h}\vec{r}]}$. Here $\vec{r}$ is a vector whose $k$ components contain a configuration of measurement results~\cite{PRXQuantum.2.030329,PhysRevA.111.023312}. At the same time, maximization is performed over all possible configurations of measurement results represented by $\vec{r}$. This is a Bell inequality for any input data $\vec{M}$ and $\tilde{C}$ compatible with the LV theory, and with any positive semi-definite matrix $A$ and vector $\vec{h}$.

We consider the measurement of $k$ local observables $\hat s_{j, \alpha}$ on a quantum system described by the density operator $\hat{\rho}$. 
Each observable has two possible outcomes $r_{j, \alpha}=\pm 1/2$ as illustrated in Fig.~\ref{fig:fig5}. 
The probability distribution $P_N(r|\alpha)$ to obtain the results $r$ given the settings $\alpha$ can be theoretically calculated, in terms of the density matrix $\hat\rho$ describing the system's state, as
\begin{align}\label{eq: DistProb}
	P_N(r|\alpha)={\rm tr}\left[\hat\rho\bigotimes_{j=1}^M\hat\Pi_{\alpha,r_{j,\alpha}}\right],
\end{align}
where $\hat\Pi_{\alpha,r_{j,\alpha}}$ projects onto the eigensubspace of $\hat s_{j, \alpha}$ with eigenvalue $r_{j,\alpha}$. 
For any particle $j$, the local measurement settings can be defined as
\begin{align}
	\label{eq:s01}
	\hat{s}_{j,\alpha} &=  \hat{s}_{j,l}\sin\theta_\alpha
	\cos\varphi_\alpha 
	+ \hat{s}_{j,l'}
	\sin\theta_\alpha\sin\varphi_\alpha
	+\hat{s}_{j,l''}\cos\theta_\alpha,
\end{align}
where $\alpha=1,2,\cdots k$ with $k$ being the number of measured observables, and $\hat{s}_{j,l}, \, \hat{s}_{j,l'},\, \hat{s}_{j,l''}$ denotes local observables with eigenvalues $\pm1/2$.
The collective observables corresponding to the measurement setting $\alpha$ for $N$ parties are given by:
\begin{align}
	\label{eq:collective settings}
	\hat{J}_{\alpha} &=  
	\hat{J}_{l}
	\sin\theta_\alpha\cos\varphi_\alpha 
	+ \hat{J}_{l'}\sin\theta_\alpha\sin\varphi_\alpha
	+\hat{J}_{l''}\cos\theta_\alpha,
\end{align}
with the collective operators defined as
$\hat{J}_{l}=\sum_{j=1}^N \hat{s}_{j,l}$,
$\hat{J}_{l'}=\sum_{j=1}^N \hat{s}_{j,l'}$,
$\hat{J}_{l''}=\sum_{j=1}^N \hat{s}_{j,l''}$.

In the case of spinor BECs, the collective operators are specified as follows: $\hat{J}_{l'', \mu}= \hat{J}^{(\sigma)}_{z,\mu}$ for all $\sigma$ and $\mu$, $\hat{J}_{l}=\hat{J}^{(s)}_{\rm{min},\mu}$ and $\hat{J}_{l'}=\hat{J}^{(a)}_{\rm{min},\mu}$ in the SU(2) subsystems $\mu = 1,2,\cdots , f$ and $\sigma = s,a$, respectively. 
The single index $\alpha$ in Eq.~(\ref{eq:collective settings}) is generalized to a pair index $(\alpha,\mu)$, with $\alpha=1,2,\cdots,k_\mu$ and $k_\mu$ being the number of setups in the given subspace $\mu$. Using this notation, the collective measurement settings then adopt the form:
\begin{align}
	\hat{J}_{\alpha,\mu} &=  
	\hat{J}^{(s)}_{\rm{min},\mu}
	\sin\theta_{\alpha,\mu}\cos\varphi_{\alpha,\mu}
	+ \hat{J}^{(a)}_{\rm{min},\mu}
	\sin\theta_{\alpha,\mu}\sin\varphi_{\alpha,\mu} \nonumber \\
	&+\hat{J}_{z,\mu}\cos\theta_{\alpha,\mu},
	\label{eq:settingschosen}
\end{align}
thus each measurement setup $(\alpha,\mu)$ corresponds to selecting a subspace $\mu$ and a direction in subspace spanned by $(\hat{J}^{(s)}_{\rm{min},\mu},\hat{J}^{(a)}_{\rm{min},\mu},\hat{J}_{z,\mu})$. The total number of possible measurements is $k=\sum_{\mu=1}^fk_\mu$.

Therefore, the components of the vector $\vec{M}$ can be written as
\begin{align}
	\frac{M_{\alpha,{\mu}}}{N} &= 
	v_\mu \cos\theta_{\alpha,\mu}
	\nonumber \\
	&+ 
	v_{\mu}^{(s)} \sin\theta_{\alpha,\mu}
	\cos\varphi_{\alpha,\mu} 
	+ v_{\mu}^{(a)} \sin\theta_{\alpha,\mu}
	\sin\varphi_{\alpha,\mu},
\end{align}
with $v_{\mu}^{(\sigma)}=\langle \hat{J}^{(\sigma)}_{\rm{min},\mu}\rangle/N$, $v_\mu=\langle \hat{J}_{z,\mu} \rangle/N$. Note that for the squeezed states we have $v_{\mu}^{(\sigma)}=0$ for all $\mu $ and $\sigma$. 
The components of the matrix $\tilde {C}$ can be written as
\begin{align}
	\label{eq:C matrix}
	&\frac{\tilde{C}^{\alpha,\mu}_{\alpha',\mu'} }{N}
	=\delta_{\mu,\mu'}\left[- \frac{1}{4}
	\cos\theta_{\alpha,\mu} 
	\cos\theta_{\alpha',\mu'}\right.  \nonumber\\
	&+
	\left(
	v_\mu ^2 \xi^2 _{\mu, s} - \frac{1}{4} \right)\sin\theta_{\alpha,\mu} \sin\theta_{\alpha',\mu'} \cos\varphi_{\alpha,\mu}
	\cos\varphi_{\alpha',\mu'} \nonumber \\
	&+ 
	\left.\left(v_\mu^2 \xi^2 _{\mu, a} - \frac{1}{4} \right)
	\sin\theta_{\alpha,\mu} \sin\theta_{\alpha',\mu'} 
	\sin\varphi_{\alpha,\mu}  \sin\varphi_{\alpha,\mu'}\right],
\end{align}
when introducing $(\Delta \hat{J}^{(\sigma)}_{\rm{min},\mu} )^2=N v_\mu^2 \xi^2_{\mu, \sigma}$
and assuming that $\hat{\rho}$ describes squeezed states for which we have
$\langle \hat{J}^{(s)}_{\rm{min},\mu} \hat{J}^{(a)}_{\rm{min},\mu} \rangle=0$,
$\langle \hat{J}^{(\sigma)}_{\rm{min},\mu} \hat{J}_{z,\mu} \rangle=0$, $(\Delta \hat{J}_{z,\mu} )^2=0$, as well as for the local measurements, respectively.
Note that the matrix $\tilde{C}$ given by Eq.~(\ref{eq:C matrix}) has a block-diagonal structure with $f$ blocks indexed by $\mu$. Each block is $k_\mu\times k_\mu$ matrix with elements indexed by $\alpha$ and $\alpha'$.

The data-driven method expressed by (\ref{eq:generalBellinequality}) is used to find the form of matrix $A$ and vector $\vec {h}$, and to construct the corresponding Bell inequality.
Numerical minimization of $L(A, \vec{h})$ for squeezed states converges to $A_{\alpha,\mu;\alpha'\mu'}=4\delta_{\mu,\mu'}w_\mu$ for all $\alpha,\mu,\alpha',\mu'$ and $h_{\alpha,\mu}=4w_\mu(2\alpha-1-k_\mu)$, where $w_\mu\ge0$ are weights between blocks. 
This gives the classical limit $E_{\rm max}=\sum_{\mu=1}^fw_\mu k_\mu^2$.
The resulting Bell inequality is additive,
\begin{align}
	\frac{{L^{(k)}_{\vec{\theta},\vec{\varphi}}}}{N}
	&=\sum_{\mu=1}^fw_\mu k_\mu^2\mathcal{L}(k_\mu,\vec{\theta}_\mu,\vec{\varphi}_\mu)\ge0,
	\label{eq:generalBE}
\end{align}
due to the block diagonal structure of the matrix $\tilde{C}$, with 
\begin{align}
	\mathcal{L}(k_\mu,\vec{\theta}_\mu,\vec{\varphi}_\mu)
	&=
	1 + k_\mu^{-2}v_\mu \sum_{\alpha=1}^{k_\mu}h_{\alpha,\mu}\cos\theta_{\alpha,\mu} \nonumber \\
	&- k_\mu^{-2}
	\left( 
	\sum_{\alpha=1}^{k_\mu}
	\cos\theta_{\alpha,{\mu}} 
	\right)^2\nonumber \\
	& - k_\mu^{-2}(1-4v_\mu^2\xi_{\mu, s}^2)
	\left(\sum_{\alpha=1}^{k_\mu} \sin\theta_{\alpha,\mu} \cos\varphi_{\alpha,\mu}\right)^2
	\nonumber \\
	&  - k_\mu^{-2}(1-4v_\mu^2\xi_{\mu, a}^2)
	\left(\sum_{\alpha=1}^{k_\mu} \sin\theta_{\alpha,\mu} \sin\varphi_{\alpha,\mu}\right)^2 .
	\label{eq:generalBE}
\end{align}
\begin{figure}[]
	\centering
	\includegraphics[width=\linewidth]{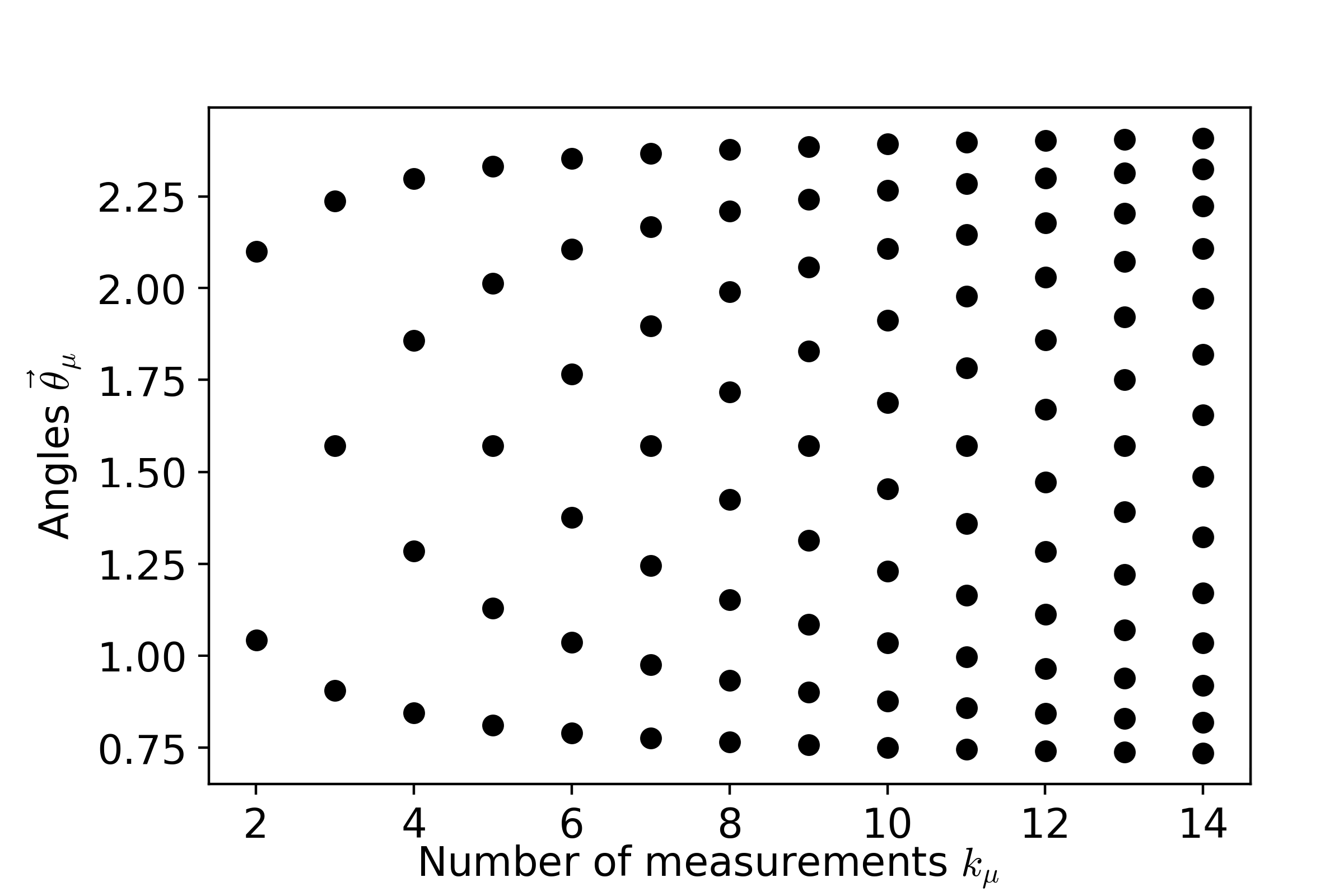}
	\caption{
		Optimal angles $\vec{\theta}$ versus the number of measurement settings $k_\mu$ as given by Eqs. (\ref{eq:thetasigma}) and (\ref{eq:thetasigma2}).
	}
	\label{fig:fig6}
\end{figure}

The minimization over angles $\vec{\theta}_{\mu},\vec{\varphi}_{\mu}$ can be performed independently within each subspace block $\mu$.
To achieve this, we assume symmetry between the symmetric and anti-symmetric subspaces, i.e.,
$\xi_\mu^2=\xi_{\mu,s}^2=\xi_{\mu,a}^2$.
Under this condition, minimization of the correlator $\mathcal{L}(k_\mu,\vec{\theta}_\mu,\vec{\varphi}_\mu)$ with respect to $\vec{\varphi}_\mu$ reveals that all values of $\varphi_{\alpha,\mu}$ are independent of $\alpha$, leading to:
\begin{align}
	\frac{\mathcal{L}(k_\mu,\vec{\theta}_\mu)}{N}
	&= \left[1 + k_\mu^{-2}v_\mu\sum_{\alpha=1}^{k_\mu}h_{\alpha,\mu}\cos\theta_{\alpha,\mu}\right.\nonumber \\
	& \left. - 
	k_\mu^{-2}\left( 
	\sum_{\alpha=1}^{k_\mu}
	\cos\theta_{\alpha,{\mu}} 
	\right)^2\right.\nonumber \\
	& \left.- k_\mu^{-2}(1-4v_\mu^2\xi_{\mu}^2)
	\left(\sum_{\alpha=1}^{k_\mu} \sin\theta_{\alpha,\mu}\right)^2\right]\ge0.
	\label{eq:generalBE-sym}
\end{align}
The minimization of $\mathcal{L}(k_\mu,\vec{\theta}_\mu)$ over angles $\vec{\theta}_\mu$ gives the following relations
\begin{align}
	\label{eq:thetasigma}
	\theta_{\alpha,\mu } &= 
	\pi - \theta_{k_\mu+1-\alpha,\mu}\\
	\label{eq:thetasigma2}
	\frac{k_\mu^2 v_\mu h_{\alpha,\mu}}{2(1-4 v_\mu^2 \xi_\mu^2)} 
	\tan{\theta_{\alpha,\mu}} &= - \sum_{\alpha' = 1}^{k_\mu} \sin{\theta_{\alpha',\mu}}.
\end{align}
In Fig.~(\ref{fig:fig6}), we present the optimized values of $\vec{\theta}_{\mu}$ versus $k_\mu$ in the thermodynamic limit with $\xi_\mu^2\to 0$ and $v_\mu=1/2$.
The angle-optimized Bell correlator takes the form of a weighted average,
\begin{equation}
	\frac{L^{(k)}_\text{opt}}{E_{\rm max}}=\frac{\sum_{\mu=1}^fw_\mu k_\mu^2 \mathcal{L}(k_\mu)}{\sum_{\mu=1}^fw_\mu k_\mu^2}\ge\underset{\mu}{\text{min}}\ \mathcal{L}(k_\mu),
	\label{eq:BelloverEmaxgeneral}
\end{equation}
with the values of $\mathcal{L}(k_\mu)$ shown in the main text. 
The final optimization over the weights $w_\mu$ selectively enhances the contribution of subsystems $\mu$ where the Bell correlation $\mathcal{L}(k_\mu)$ exhibits significant negativity, while assigning zero weight to subsystems where the negativity of $\mathcal{L}(k_\mu)$ is negligible or absent.
On the other hand, if $k_\mu^2 \mathcal{L}(k_\mu)$ is negative and varies only weakly with $\mu$, it is beneficial to use the arithmetic mean by setting $w_\mu=1/f$, thus increasing contribution from all subsystems, leading to a larger total number of settings $k$.

It should be noted that the formula for the multisetting two-body Bell correlator presented in~\cite{PhysRevLett.119.170403} is a special case of (\ref{eq:generalBE-sym}) within a given subsystem $\mu$ and assuming the symmetry between $a$ and $s$.

\section{Spin-1 BECs}
\label{sec:Spin-1 BECs.-}

This section provides a step-by-step application of the theory introduced in Sec.~\ref{sec:Bellinequality} and Appendix~\ref{app:Bellscenario} for the case of spin $f=1$ BEC. We consider a system of rubidium-87 atoms prepared in the $F=1$ ground hyperfine manifold. Under the SMA, the system is described by the interaction Hamiltonian, Eq.(1) of the mains text, with $c_2=c_3=0$, which can be simplified as following:
\begin{align}
	\hat{H}_{\rm int}
	&=
	\frac{c_1}{2N }
	\left[
	2 (\hat{a}_1^\dagger \hat{a}_{-1} ^\dagger \hat{a}_0\hat{a}_{0} +h.c.) 
	+(2\hat{N}_0-1)(\hat{N} - \hat{N}_0)
	\right],
\end{align}
where $\hat{a}_m$ is an anihilation operator of an atom in the state $|F=1, m=0,\pm1\rangle $.
The Hamiltonian above includes an interaction term that drives spin-mixing processes by generating a pair of atoms in the $|F=1, m=\pm 1\rangle $ states from the collision of two atoms in the $|F=1, m=0\rangle $ state, and vice versa.
This interaction generates spin-nematic squeezed states in the symmetric and antisymmetric subspaces defined in the main text for $\mu=1$, simultaneously, 
when the initial state is prepared with all atoms in the hyperfine state $|F=1, m=0,\pm1\rangle $. 

The squeezing parameter in the symmetric and antisymmetric subspaces, $\xi^2_{1, s}$ and  $\xi^2_{1, a}$, were measured simultaneously in recent experiment~\cite{cao2023jointestimationtwophasespin} by using atomic homodyne detection method~\cite{Gross_2011}. The metrological gains observed for join estimation of the two phases reach $3.3$dB to $6.3$dB beyond the standard quantum limit over a wide range of parameters~\cite{cao2023jointestimationtwophasespin}. The technical advantage of the simultaneous measurement in two subspaces makes the spin-1 system an ideal platform for testing the multisetting Bell scenario introduced in Sec.~\ref{sec:results}. 

We begin the analysis by recalling the local measurement setting (\ref{eq:s01}), which leads to collective measurements performed in the symmetric and antisymmetric subspaces, along with the corresponding chosen settings~(\ref{eq:settingschosen}), as follows:
\begin{align}
	\label{eq:collective settings f=1}
	\hat{J}_{\alpha, 1} &=  \hat{J}^{(s)}_{\rm{min},1} \sin(\theta_{\alpha, 1})\cos(\varphi_{\alpha, 1}) 
	+ \hat{J}^{(a)}_{\rm{min},1}\sin(\theta_{\alpha, 1})\sin(\varphi_{\alpha, 1})
	\nonumber \\
	&+\hat{J}_{z, 1}\cos(\theta_{\alpha, 1}),
\end{align}
where $\alpha=1,2,3$ and $k_\mu=3$. 
The vectors $\hat{J}^{(\sigma)}_{\rm{min},1}$ for $\sigma = s,a$ are given by the best squeezing direction in the symmetric and anti-symmetric subspaces, respectively. 

The general Bell inequality (\ref{eq:generalBellinequality}) reduces to:
\begin{align}
	L^{(3)}_{\vec{\theta},\vec{\varphi}} & = 9N + 8(M_0-M_2) \nonumber \\
	&+4
	\left(\tilde{C}_{0,1}^{0,1}+\tilde{C}_{1,1}^{1,1}+\tilde{C}_{2,1}^{2,1} +2 \tilde{C}_{0,1}^{1,1}+2 \tilde{C}_{0,1}^{2,1}+2 \tilde{C}_{1,1}^{2,1}\right)\geq 0
	\label{eq:21}
\end{align}
with
\begin{equation}
	A=
	4\begin{bmatrix}
		1 & 1 & 1 \\
		1&1&1\\
		1 & 1 &1
	\end{bmatrix},
	\,\,\, 
	h^T=(8,0,-8),
\end{equation}
(see Section~\ref{app:Bellscenario} for derivation).
We express $L^{(3)}_{\vec{\theta},\vec{\varphi}}$ in terms of the normalized spin squeezing parameter 
$\xi_s^2=N (\Delta\hat{J}^{(s)}_{\rm{min},1})^2 /\langle\hat{J}^{(s)}_{z, 1}\rangle^2$, 
$\xi_a^2=N (\Delta\hat{J}^{(a)}_{\rm{min},1})^2/\langle\hat{J}^{(s)}_{z, 1}\rangle^2$ and normalized mean spin $v=\langle\hat{J}_{z, 1}\rangle/N$,
as:
\begin{align}
	\frac{L^{(3)}_{\vec{\theta},\vec{\varphi} } }{E_{\rm max}}
	&= 1 + 8/9(v \cos\theta_0 -  v \cos\theta_2) \nonumber \\
	& - 9^{-1} (1-4v^2\xi_s^2)\left(\sum_{\alpha=1}^k \sin\theta_\alpha \cos\varphi_\alpha\right)^2\nonumber \\
	& - 9^{-1} (1-4v^2\xi_a^2)\left(\sum_{\alpha=1}^k \sin\theta_\alpha\sin\varphi_\alpha\right)^2.
	\label{eq:thetavarphiN}
\end{align}
This is Eq. (\ref{eq:generalBE}) for $\mu = 1$ and $k_\mu = 3$. Minimization concerning $\varphi_\alpha$ gives two sets of solutions:
$\cos\varphi_\alpha=1$ if $\xi_s^2<\xi_a^2$,
and
$\cos\varphi_\alpha=0$ if $\xi_a^2<\xi_s^2$.
This ensures that the contribution from the terms involving a larger squeezing parameter is suppressed, thereby enhancing the Bell inequality violation.

We now consider the symmetric case, where $\xi_s^2=\xi^2_a=\xi^2$, and focus on minimizing the Bell correlator~(\ref{eq:thetavarphiN}) with respect to the angles $\theta_{1}$, $\theta_{2}$ and $\theta_{3}$. In the large $N$ limit, this yields the optimal conditions:
\begin{align}
	\theta_{2\, {\rm opt}} &= \frac{\pi}{2} 
	\label{eq:theta0opt} \\
	\theta_{1\, {\rm opt}} &= \pi - \theta_{3\, {\rm opt}} \\
	\frac{4 v}{1-4 v^2 \xi^2} \tan \theta_{3\, {\rm opt}} &= 1 + 2 \sin \theta_{3\, {\rm opt}} ,\label{eq:theta2opt}
\end{align}
leading to the optimal Bell inequality
\begin{equation}
	\label{eq:violation for k=3}
	\frac{L^{(3)}_{{\rm opt}}}{E_{\rm max} } =1-\frac{16}{9}v\cos\theta_{3\text{ opt}}-\frac{1-4v^2\xi^2}{9}(1+2\sin\theta_{3\text{ opt}})^2.
\end{equation}
In general, one needs to calculate the value of $\theta_{3,{\rm opt}}$ from (\ref{eq:theta2opt}) for given $v,\, \xi^2$, and insert it into (\ref{eq:violation for k=3}) to obtain the Bell correlator.
In the thermodynamics limit, however, we have $v=1/2$, $\xi^2 \to 0$, which leads to $\theta_{3\, {\rm opt}}=\arctan(\sqrt{5/3}) \approx 0.91$. Substituting these values, we find
\begin{equation}
	\label{eq:lowerband}
	\frac{L^{(3)}_{\rm opt}}{E_{\rm max}} \bigg|_{N\to \infty} \approx -0.285.
\end{equation}
This value represents a $12\%$ stronger violation compared to the two-measurement Bell scenario. This demonstrates enhanced Bell detection with three measurement settings.
In Fig.~2 of the main text we present the Bell correlator ${L^{(3)}_{{\rm opt}}  /E_{\rm max}}$ as a function of time for $q=|c_1|$, using experimental data from \cite{cao2023jointestimationtwophasespin}. 
The experimental data points for the Bell correlator defined in Eq.~(\ref{eq:thetavarphiN}), with angles optimized analytically as per Eqs.~(\ref{eq:theta0opt})--(\ref{eq:theta2opt}), are represented there by blue dots.

It is worth noticing that when additional correlations are introduced between the symmetric and anti-symmetric operators, such that $\langle \hat{J}_{\text{min},1}^{(s)} \hat{J}_{\text{min},1}^{(a)} \rangle \ne 0$, the value of the Bell correlator decreases even more. This leads to the violation of the bound given in Eq.(\ref{eq:lowerband}). The processes that could generate such correlations are not naturally present in the spin-1 Hamiltonian. More detailed discussions of this effect can be found in Section~\ref{app:mixing}. 
We also analyze the effect of imperfections with states $\hat{\rho} = p\, \hat{\rho}_S + (1-p)\, \hat{\rho}_\perp$, where $ \hat{\rho}_\perp$ represents either a coherent state $ = |\psi_0\rangle \langle \psi_0|$ or a maximally mixed state $ \hat{\rho}_\perp = \mathbb{I}/\mathcal{N}$, with further details provided in Section~\ref{app:mixedstates}.

\section{Spin-2 and spin-3 BECs}
\label{appsec:Spin-2 and spin-3 Becs}

In this section, we examine the cases of $f=2$ and $f=3$, corresponding to atoms such as rubidium or sodium ($f=2$, Zeeman levels $m=0,\pm1,\pm2$) and chromium ($f=3$, $m=0,\pm1,\pm2,\pm3$). We assume the dynamics are governed by the Hamiltonian (\ref{eq:ham}), with $c_1, c_2 \neq 0$, and $c_3 = 0$ for $f=2$, while for $f=3$, all three coefficients $c_1, c_2,$ and $c_3$ are non-zero. For $f=2$, the system exhibits two pairs of symmetric and antisymmetric subspaces, whereas for $f=3$, there are three such pairs, as defined in the main text.

\begin{figure}[]
	\centering
	\includegraphics[width=\linewidth]{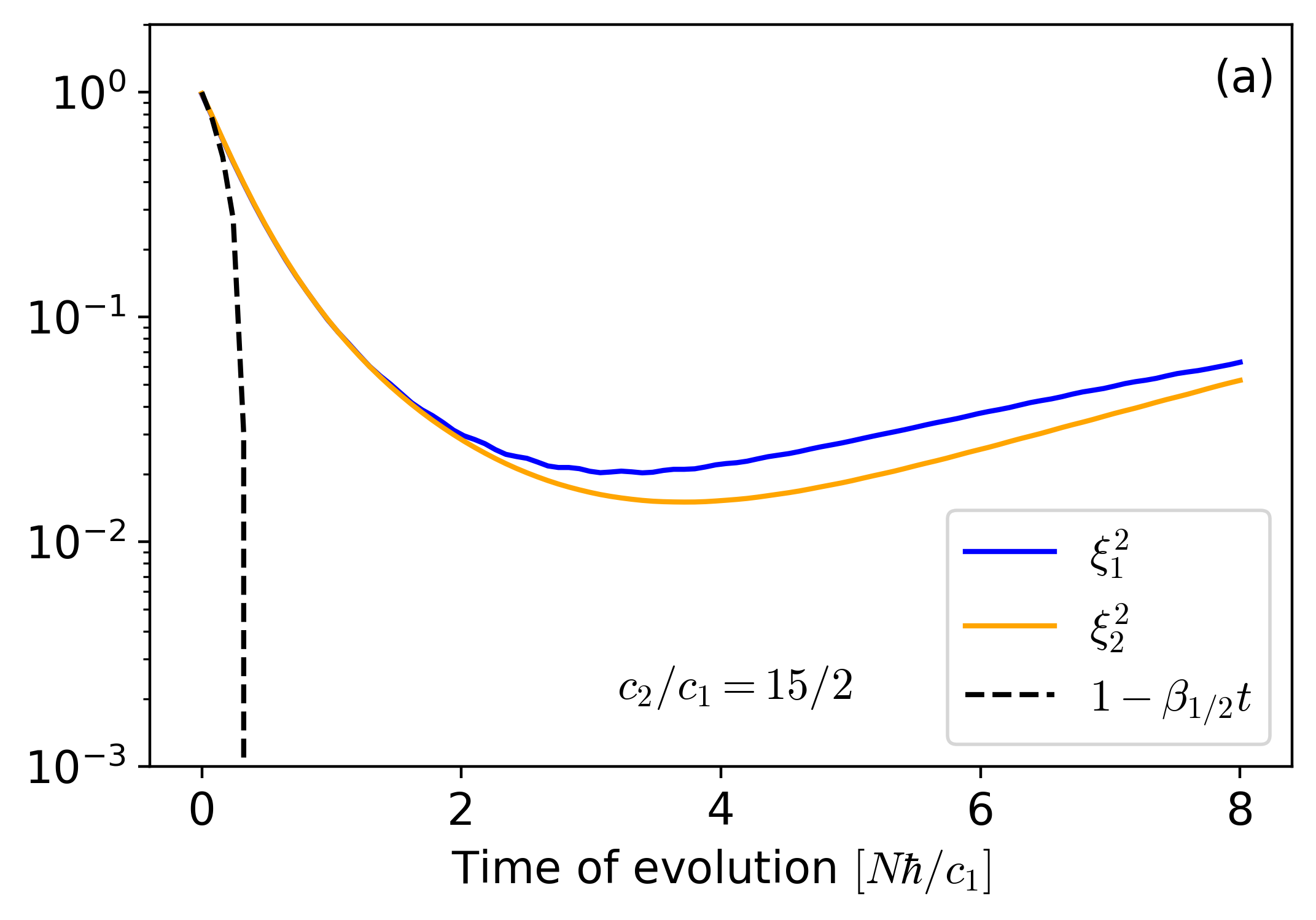}
	\includegraphics[width=\linewidth]{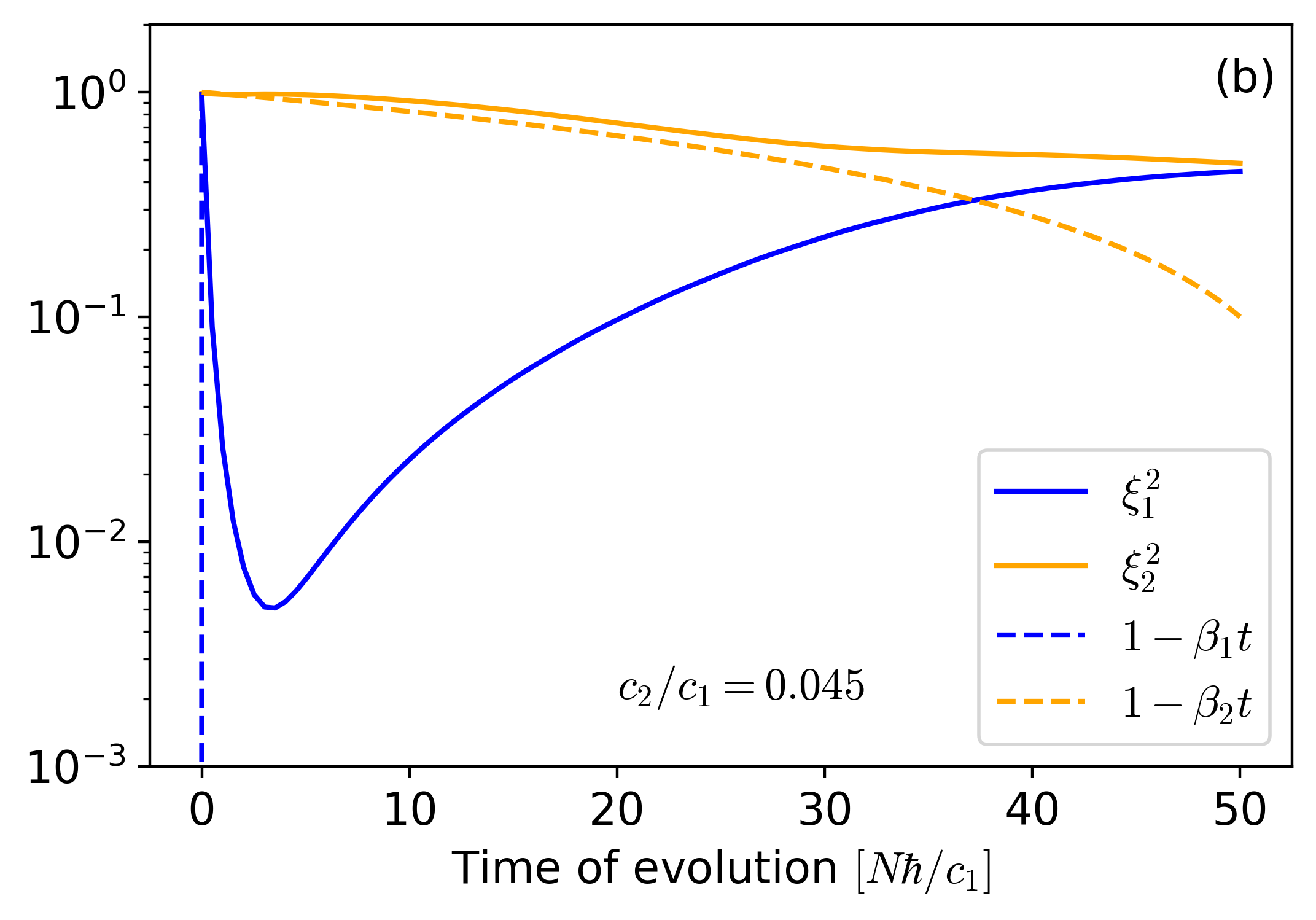}
	\caption{
		The dynamics of the squeezing parameter for $f=2$ in the $\mu=1$ (blue solid line) and $\mu=2$ (orange line) subsystems for spin-2 BECs for $c_2/c_1=15/2$ (a) and $c_2/c_1=0.045$ (b) when $N=26400$ and $10^5$ realizations in the TWM. The dashed lines indicate the squeezing rate given by (\ref{eq:squeezveryshorttimes}). 
		}
	\label{fig:fig3}
\end{figure}

\begin{figure}[]
	\centering
	\includegraphics[width=\linewidth]{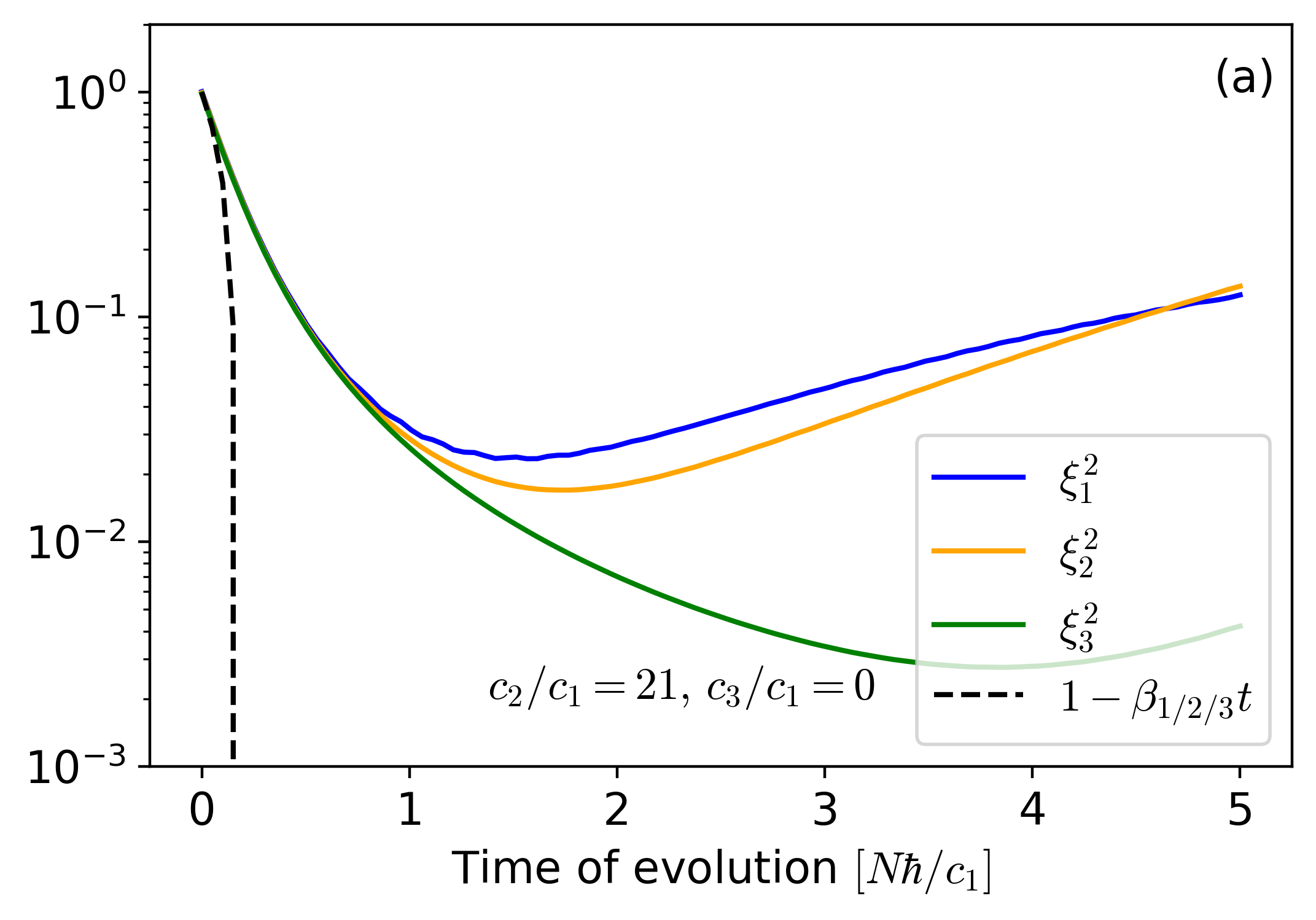}
	\includegraphics[width=\linewidth]{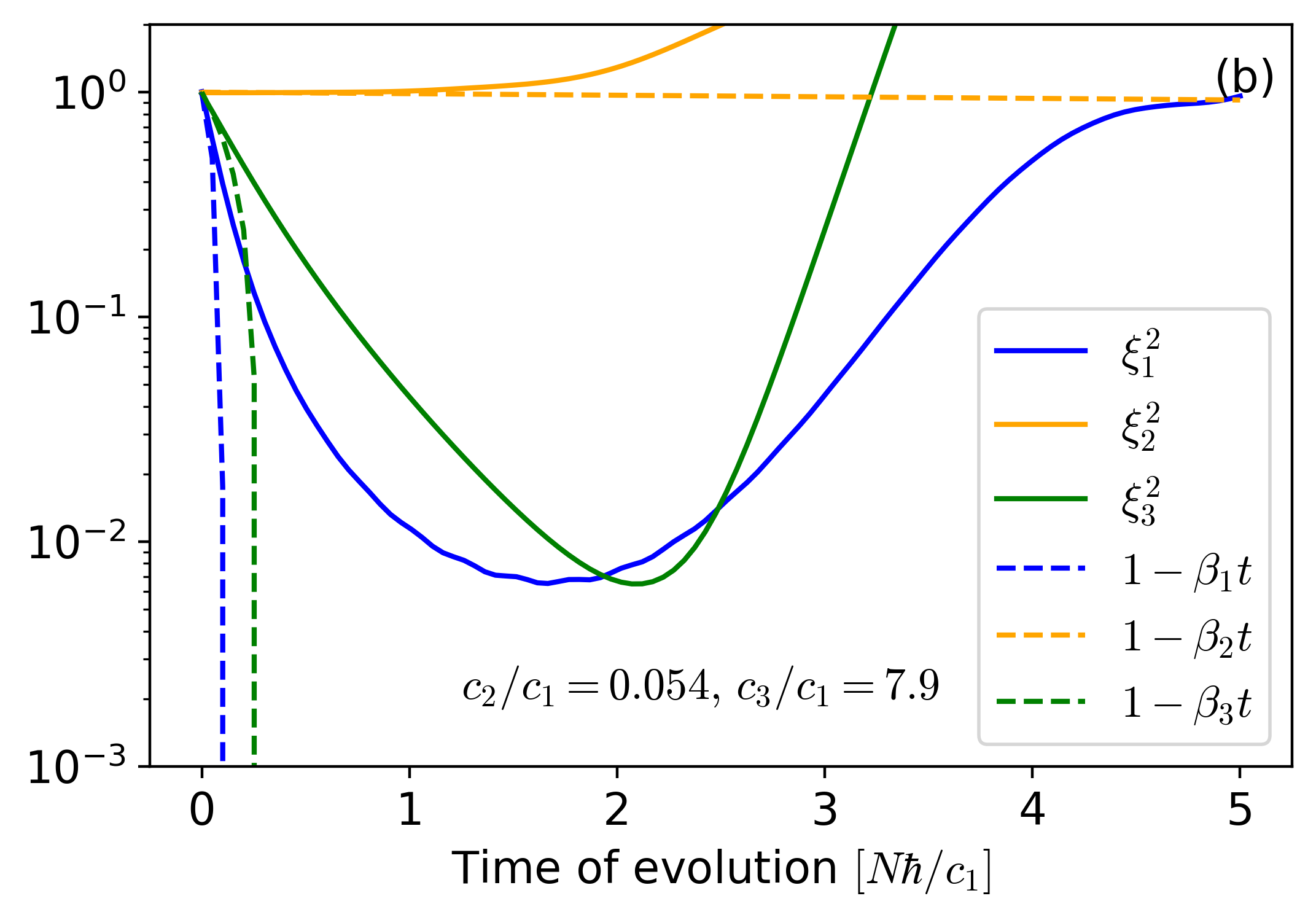}
	\caption{The squeezing parameter for $f=3$ in the $\mu=1$ (blue solid line), $\mu=2$ (orange) and $\mu=3 $ (green) subsystems with parameters $c_2=105c_1/4,\,c_3=63c_1/2$ and $N=10$ is shown in panel (a). Panel (b) presents the case with $\,c_2=420c_1/11,\,c_3=6/5c_2=504c_1/11$ and $N=10$. The dashed lines indicate the initial squeezing rate given by (\ref{eq:squeezveryshorttimes}).
	}
	\label{fig:fig4}
\end{figure}

In Figs.~\ref{fig:fig3} and~\ref{fig:fig4}, we illustrate the evolution of the squeezing parameter within the corresponding subsystems. The generated squeezing level shows a dependence on the parameter $\mu$. This dependence originates from the interaction terms in the Hamiltonian proportional to $c_2$ and $c_3$. These terms introduce additional interaction channels, specifically $\hat{a}_0^2 \hat{a}_\mu^\dagger \hat{a}_{-\mu}^\dagger + \text{h.c.}$ for $\mu = 2, 3$, which affect the rate of squeezing generation in each subsystem. We present two distinct scenarios: (a) when the initial rate of squeezing generation is uniform across specific subsystems, and (b) when squeezing is dominant or negligible in one of the subsystems.
These scenarios result in different levels of Bell inequality violation.

We first calculate the initial rate of squeezing generation by analyzing the nontrivial terms in the short-time expansion of the dynamics:
\begin{align}
	\xi^2_\mu&=
	1 - \beta_\mu t + \mathcal{O}(t^2),
	\label{eq:squeezveryshorttimes}
\end{align}
where we introduce the initial slope $\beta_\mu$ governing the variation of $\xi^2_\mu$, defined as
\begin{equation}
	\beta_\mu 
	= \frac{d\xi^2_\mu}{dt}|_{t=0}=
	\frac{4}{N\hbar}
	|
	\langle \psi_0| 
	i
	[ \hat{H}, (\Delta \hat{J}^{(\sigma)}_{\rm{min},\mu})^2 ]
	|\psi_0 \rangle
	|
	\label{eq:analytical strength}
\end{equation}
with the expectation value evaluated over the initial state $|\psi_0\rangle$, see main text for definitions.
Given that
$\hat{J}^{(\sigma)}_{\rm{min},\mu} = \cos\phi\hat{J}_{x,\mu}^{(\sigma) } + \sin\phi\hat{J}_{y,\mu}^{(\sigma)}$, with $\phi$ being the best squeezing direction, we have 
\begin{align}
	&\beta_\mu=
	\frac{4}{N\hbar}
	\left(
	\begin{matrix}
		\cos\phi, & \sin\phi \end{matrix}
	\right)
	\mathcal{M}_\mu^{(\sigma)}(\hat{H})
	\left(
	\begin{matrix}
		\cos\phi\\\sin\phi
	\end{matrix}
	\right).
\end{align}
where is the matrix of commutators:
\begin{equation}
	\mathcal{M}_\mu^{(\sigma)} = 
	\begin{bmatrix}
		\langle \psi_0| i[\hat{H},(\hat{J}_{x,\mu}^{(\sigma)})^2]|\psi_0\rangle &
		\langle \psi_0| i[\hat{H},\hat{J}_{x,\mu}^{(\sigma)}\hat{J}_{y,\mu}^{(\sigma)}]|\psi_0 \rangle\\
		\langle \psi_0| i[\hat{H},\hat{J}_{y,\mu}^{(\sigma)}\hat{J}_{x,\mu}^{(\sigma)}] |\psi_0 \rangle &
		\langle \psi_0| i[\hat{H},(\hat{J}_{y,\mu}^{(\sigma)})^2] |\psi_0 \rangle 
		\nonumber
	\end{bmatrix}.
\end{equation}

We derive analytically the expectation values of commutators for all terms in the interaction Hamiltonian $\hat{H}_{\rm int}$, treating each term independently.
Nonzero contributions originate solely from spin-exchange interactions of the form $\hat{H_\mu}=\hat{a}_0^2 \hat{a}_\mu^\dagger\hat{a}_{-\mu}^\dagger+$ h.c., within magnetization-conserving channels, $\mathcal{M}=0$.
This result arises due to the initial state we choose comprising particles solely in the $m=0$ state.
For the spin-exchange terms in the interaction Hamiltonian, we derive the following elements of the $\mathcal{M}_\mu^{(\sigma)}$ matrix:
\begin{equation}
	\mathcal{M}_\mu^{(\sigma)}(\hat{H}=\hat{H_\mu}) =-\frac{N^2}{2}\begin{bmatrix}
		0&1\\1&0
	\end{bmatrix},
\end{equation}
independent of $\sigma = s,a$ and $\mu=1,2,3$.
Summing contributions from all terms in the interaction Hamiltonian, we obtain
\begin{equation}
	\hbar \beta_1 = 2 \, \left| c_1\right| N,
\end{equation}
for $f=1$,
\begin{align}
	\hbar \beta_1& = \left| 6 -  \frac{2c_2}{5c_1} \right| \, |c_1|  N,\\
	\hbar \beta_2& = \left| \frac{2 c_2 }{5 c_1 } \right| \,  |c_1 | N ,
\end{align}
for $f=2$, and 
\begin{align}
	\hbar \beta_1& = \left|12 - \frac{2c_2}{7c_1} - \frac{2c_3}{7c_1}\right|\, |c_1| N,\\
	\hbar \beta_2& = \left| \frac{2 c_2 }{7 c_1 } \right| \,  |c_1| N,\\
	\hbar \beta_3& = \left|\frac{2c_2}{7c_1} -  \frac{10c_3}{21c_1}\right| \,  |c_1|N,
\end{align}
for $f=3$, when $\phi=\pm\pi/4$. 
Additionally, one can perform similar calculations for the quadratic Zeeman term by considering $\mathcal{M}_\mu^{(\sigma)}(\hat{H})$ when $\hat{H}=q\hat{N}_0$. 
This term yields no contribution when evaluating expectation values in the initial state $|\psi_0\rangle$. However, over longer timescales, 
this term significantly enhances the rate of optimal squeezing generation—an effect unaccounted for by the short-time expansion of squeezing dynamics.
In Figs.~\ref{fig:fig3} -\ref{fig:fig4} we plot the initial slopes $\beta_\mu$ with dashed lines, demonstrating the validity of the above analysis.

The simple analytical expressions derived above enable us to establish relationships between the parameters $c_1,\, c_2,\, c_3$ that produce target $\beta_\mu$. 
For instance, requiring equal initial squeezing rates across all subsystems $\mu$ yields 
$c_1=0,\, 4c_2/30$ and any $c_2$ for $f=2$, 
$c_3=0$, 
$c_1=0, \, c_2/21$ and $c_3=12c_2/10$, 
$c_1=c_2/35, 8c_2/105$ and any $c_2$ for $f=3$.
These parameters, which produce uniform initial decay rates of squeezing parameters, are shown in Figs.~\ref{fig:fig3}(a) and~\ref{fig:fig4}(a). Such results underscore the necessity of fine-tuning coupling coefficients to balance contributions from distinct interaction channels during the system’s early evolution.

On the other hand, the natural values of the scattering length for $f=2$ sodium-23 and rubidium-87 give $|c_2/c_1|\approx 1.57$ and $|c_2/c_1|\approx 0.045$~\cite{KAWAGUCHI2012253, Widera_2006}, respectively, leading to the type of scenario illustrated in panel (b) of Fig.~\ref{fig:fig3} where significant level of squeezing is generated only in the $\mu=1$ subsystem
~\footnote{The scattering length taken are 
	$a_0\approx 34.9 a_B$, $a_2\approx 45.8 a_B$ and $a_4\approx 62.51 a_B$ for sodium-23 atoms~\cite{KAWAGUCHI2012253}; 
	and
	$a_0\approx 87.93 a_B$, $a_2\approx 91.28 a_B$ and $a_4\approx 98.98 a_B$ for rubidium-87 atoms~\cite{Widera_2006}, both with spin $f=2$. Here $a_B$ is the Bohr radius.}.
For $f=3$ chromium atoms, we have
$|c_2/c_1| \approx 0.054$ and $|c_3/c_1| \approx 7.9$~\cite{Chomaz_2023}. In this case, the scenario where the squeezing level is non-negligible in both the $\mu=1$ and $\mu=3$ subsystems is more likely, it is for scattering length values 
	$a_0\approx 13.5 a_B$, $a_2\approx -7 a_B$, $a_4\approx 56 a_B$ and $a_6\approx 102.5 a_B$ 
	taken from~\cite{Chomaz_2023},
, as illustrated in Fig.~\ref{fig:fig4} (b).

The dynamics of the corresponding Bell correlator when using the squeezing parameter presented in Figs.~\ref{fig:fig3} and~\ref{fig:fig4} is discussed in the main text.

\section{Mixing symmetric and anti-symmetric subspaces in spin-$f$ systems}
\label{app:mixing}

In the case of spin-$f$ BECs, spin correlations in the symmetric and anti-symmetric subspaces can be measured through the covariance between the operators $\hat{J}^{(s)}_{\rm{min},\mu}$ and $\hat{J}^{(a)}_{\rm{min},\mu}$, specifically $\langle \{\hat{J}^{(s)}_{\rm{min},\mu} , \hat{J}^{(a)}_{\rm{min},\mu}\} \rangle$.
Such correlations are absent in systems governed by the Hamiltonian for $f=1,\,2,\, 3$. 

To derive an interaction Hamiltonian conducive to analyzing these correlations, we focus on the leading contribution to the time evolution of the covariance:
\begin{align}
	&\frac{d}{dt}\langle\psi_0|e^{i \hat{H}t/\hbar} \{ \hat{J}^{(s)}_{\rm{min},\mu} , \hat{J}^{(a)}_{\rm{min},\mu} \} e^{-i \hat{H}t/\hbar}|\psi_0\rangle=\nonumber \\
	&    =\frac{1}{\hbar}\langle\psi_0|i[\hat{H}, \{ \hat{J}^{(s)}_{\rm{min},\mu} , \hat{J}^{(a)}_{\rm{min},\mu} \} ]|\psi_0\rangle + \mathcal{O}(t),
	\label{eq:A1}
\end{align}
which is the first-order term (evaluated at $t=0$), as it is the most significant on the short timescale of squeezing. Higher-order contributions involve correlations induced by multiple Hamiltonian terms but are neglected here.
This approach yields an interaction Hamiltonian that
provides the interaction Hamiltonian that produces nonzero covariance while conserving total magnetization:
\begin{equation}
	\frac{\hat{H}_\text{mix}}{|c_1|}
	=\frac{\varepsilon}{N}
	\sum_\mu 
	\left[
	(\hat{a}_0^\dagger)^2 \hat{g}_\mu^{(s)}\hat{g}_\mu^{(a)} 
	+ \hat{g}_\mu^{{(s)}\dagger} \hat{g}_\mu^{(a)\dagger} \hat{a}_0^2
	\right].
	\label{eq:mixing}
\end{equation}
It introduces additional correlations between symmetric and anti-symmetric subspaces in the given subsystem $\mu$ without significantly affecting the generation of spin squeezing.

\begin{figure}[]
	\centering
	\includegraphics[width=\linewidth]{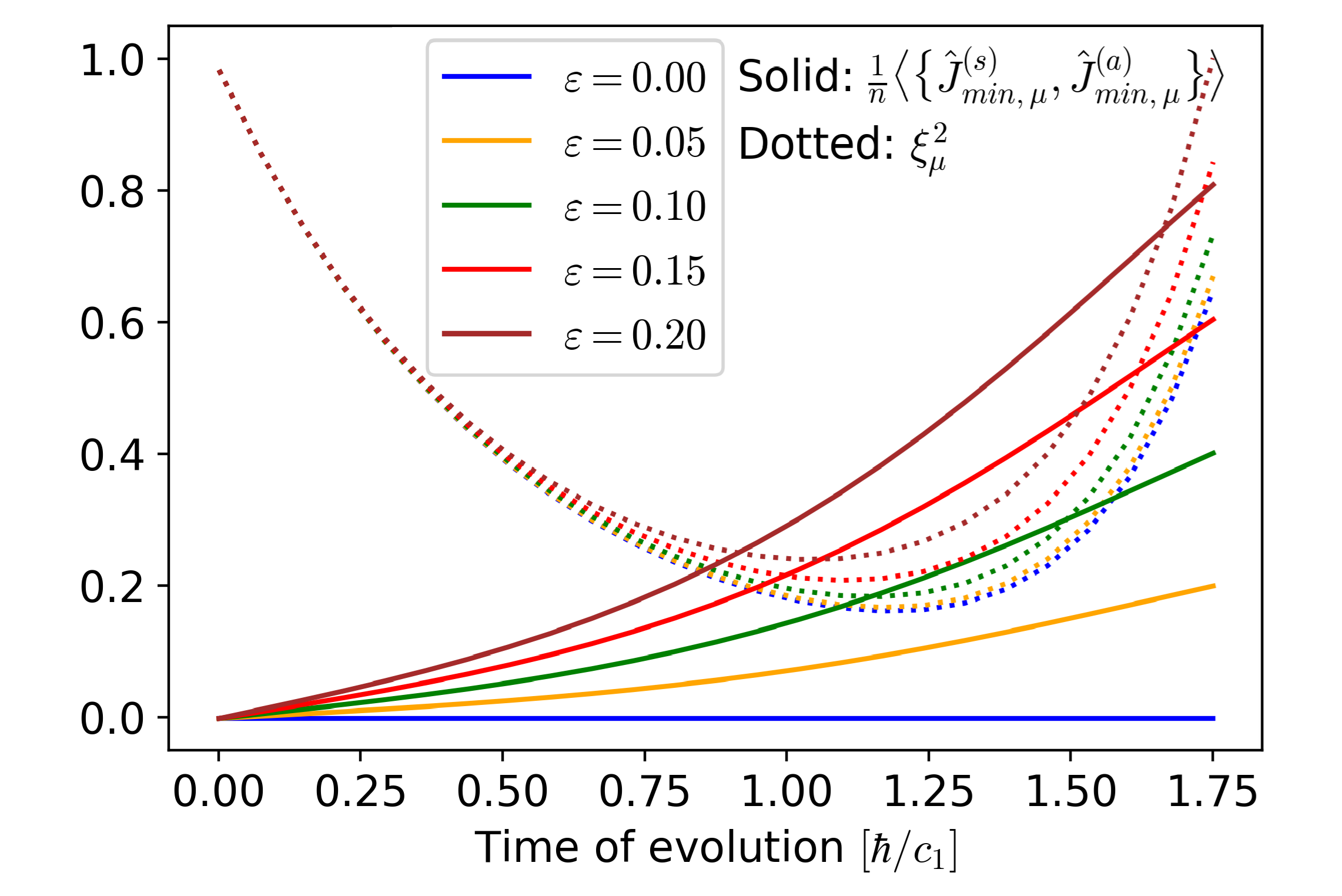}
	\caption{The squeezing parameter $\xi^2_\mu$ (dotted lines) and the anticommutator $\langle \{\hat{J}^{(s)}_{\rm{min},\mu} , \hat{J}^{(a)}_{\rm{min},\mu}\} \rangle$ (solid lines) as a function of time for several choices of $\varepsilon$ when $N=50,q=1$.}
	\label{fig:mixing term time dependence}
\end{figure}
For simplicity of the analysis, we focus on the $f=1$ case, $c_2=c_3 = 0$ in the BEC Hamiltonian, so the full Hamiltonian reads:
\begin{equation}
	\label{eq:mixing hamiltonian}
	\frac{\hat{H}_{\rm S}}{|c_1|} = 
	-\frac{1}{2N} \hat{J}^2 
	+ q \hat{N}_s 
	- \frac{\hat{H}_\text{mix}}{|c_1|}.
\end{equation}
Figure~\ref{fig:mixing term time dependence} illustrates the evolution of the anticommutator
$\langle \{\hat{J}^{(s)}_{\rm{min},\mu} , \hat{J}^{(a)}_{\rm{min},\mu}\} \rangle$
and the squeezing parameter as a function of $\varepsilon$.
The non-zero mixing term $\varepsilon$ slightly increases the squeezing parameter. However, since the growth of $\xi^2$ is quadratic, a sufficiently small value of $\varepsilon$ induces the required inter-subspace correlations while preserving near-optimal squeezing.

The Bell correlator, defined by the left-hand side of (\ref{eq:generalBellinequality}), can be employed with a modified covariance matrix $\tilde{C}$. 
We choose the measurement settings as in Eq.~(\ref{eq:s01}), which results in the following non-zero elements of the covariance matrix: 
\begin{align}
	\tilde{C}^{\alpha, \mu}_{\alpha' \mu} 
	&= \left[ (\Delta\hat{J}_{z, \mu})^2 - \frac{N}{4}
	\right]\cos(\theta_\alpha) \cos(\theta_{\alpha'})  \nonumber\\
	&+
	\left[(\Delta \hat{J}^{(s)}_{\rm{min},\mu})^2-\frac{N}{4}
	\right]\sin(\theta_{\alpha}) \sin(\theta_{\alpha'}) \cos(\varphi_\alpha)\cos(\varphi_{\alpha'}) \nonumber \\
	&+ 
	\left[(\Delta\hat{J}^{(a)}_{\rm{min},\mu})^2-\frac{N}{4}
	\right]\sin(\theta_{\alpha}) \sin(\theta_{\alpha'}) \sin(\varphi_\alpha)  \sin(\varphi_{\alpha'})
	\nonumber \\
	& + \left[\langle \{\hat{J}^{(s)}_{\rm{min},\mu}, \hat{J}^{(a)}_{\rm{min},\mu} \} \rangle 
	-\sum_j
	\langle \{  \hat{s}^{(s)}_{{\rm min},\mu, j}  , \hat{s}^{(a)}_{{\rm min},\mu, j}  \} \rangle\right] \times \nonumber \\
	& 
	\times 
	\sin(\theta_\alpha) \sin(\theta_\alpha') 
	\sin(\varphi_\alpha) \cos(\varphi_{\alpha'}),
\end{align}
where we neglect the contributions from
$\langle \hat{J}^{(s)}_{\rm{min},\mu} \hat{J}^{(\sigma)}_{z,\mu} \rangle$, 
$\langle \hat{J}^{(a)}_{\rm{min},\mu} \hat{J}^{(\sigma)}_{z,\mu} \rangle$,
due to their negligible magnitudes.
Next, we optimize the Bell correlator over the angles $\varphi_\alpha, \theta_\alpha$ in the symmetric case, where the squeezing parameter is identical in both symmetric and antisymmetric subspaces.
In the thermodynamic limit, we 
obtained $\varphi_\alpha=\pi/4$ and $\theta_\alpha$ given by (\ref{eq:thetasigma}),(\ref{eq:thetasigma2}). 
Finally, we recover the same Bell inequality as in the $\varepsilon=0$ case, given by Eq.~(\ref{eq:generalBE}), with the crucial distinction that the $(1-4 v^2 \xi^2)$
term acquires an additional $(1-\eta)$ factor. Here $\eta$ is defined as
\begin{equation}
	\eta = \frac{\langle \{ \hat{J}^{(s)}_{\rm{min},\mu}, \hat{J}^{(a)}_{\rm{min},\mu} \} \rangle 
		-N\langle \{  \hat{s}^{(s)}_{{\rm min},\mu,j} , \hat{s}^{(a)}_{{\rm min},\mu,j} \} \rangle }{(\Delta\hat{J}^{(a)}_{\rm{min},\mu})^2-\frac{N}{4}.
	}
\end{equation}
The modified Bell correlator then becomes:
\begin{align}
	\frac{L^{(3)}_{{\rm opt}}(\eta)}
	{E_{\rm max}}
	&=1-\frac{16}{9}v\cos\theta_{3\, \rm{opt}} \nonumber \\
	&-(1+\eta)\frac{1-4v^2\xi^2}{9}(1+2\sin\theta_{3\,\rm{opt}})^2.
\end{align}
In the thermodynamic limit $N\to \infty$, where $\xi^2\to 0$ and $v\to 1/2$, this simplifies to:
\begin{align}
	\frac{L^{(3)}_{{\rm opt}}(\eta)}
	{E_{\rm max}} \bigg | _{N\to \infty}
	&\approx 0.4556 - (1+\eta) \, 0.7402 .
\end{align}
The maximal value of $\eta$ is $\eta=1$, when $\langle \{ \hat{J}^{(s)}_{\rm{min},\mu} , \hat{J}^{(a)}_{\rm{min},\mu} \} \rangle \ll N$ and $(\Delta\hat{J}^{(a)}_{\rm{min},\mu})^2\ll N$, which gives 
$
L^{(3)}_{{\rm opt}}(\eta)
/ E_{\rm max}
\approx -1.0248$.
This indicates that correlations generated by the mixing Hamiltonian (Eq.~\ref{eq:mixing hamiltonian}) enhance Bell correlations in this regime.

\begin{figure}[]
	\centering
	\includegraphics[width=\linewidth]{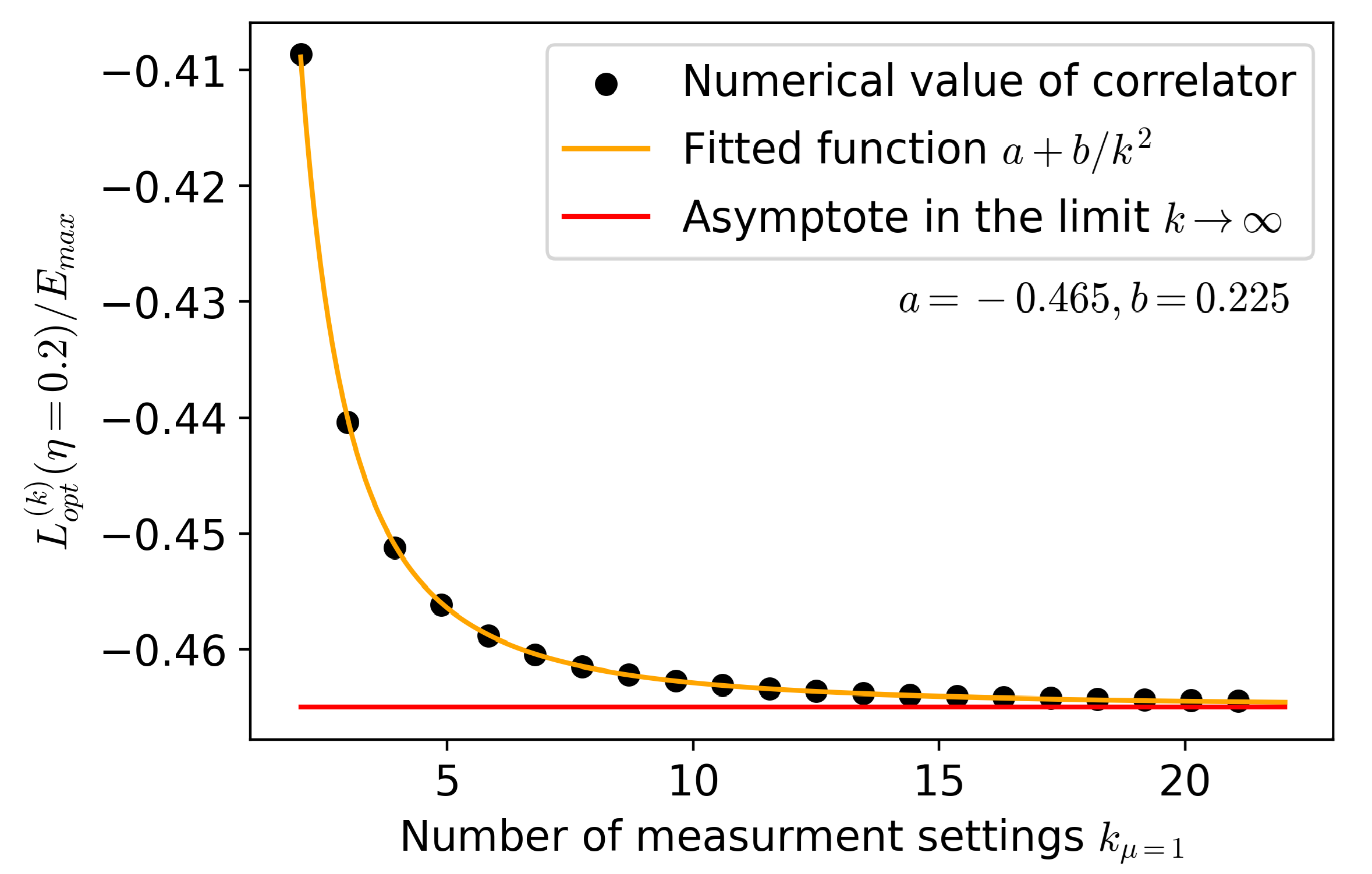}
	\caption{The minimal value of the Bell correlator $L^{(k)}_{{\rm opt}}(\eta=0.2)/E_{\rm max}$ for $f=1$ and $\eta=0.2$ versus the number of measurement settings $k_{\mu=1}$. 
		Inset shows the asymptotic value of the Bell correlator versus $\eta$.}
	\label{fig:asymptotic}
\end{figure}

\begin{figure}[]
	\centering
	\includegraphics[width=\linewidth]{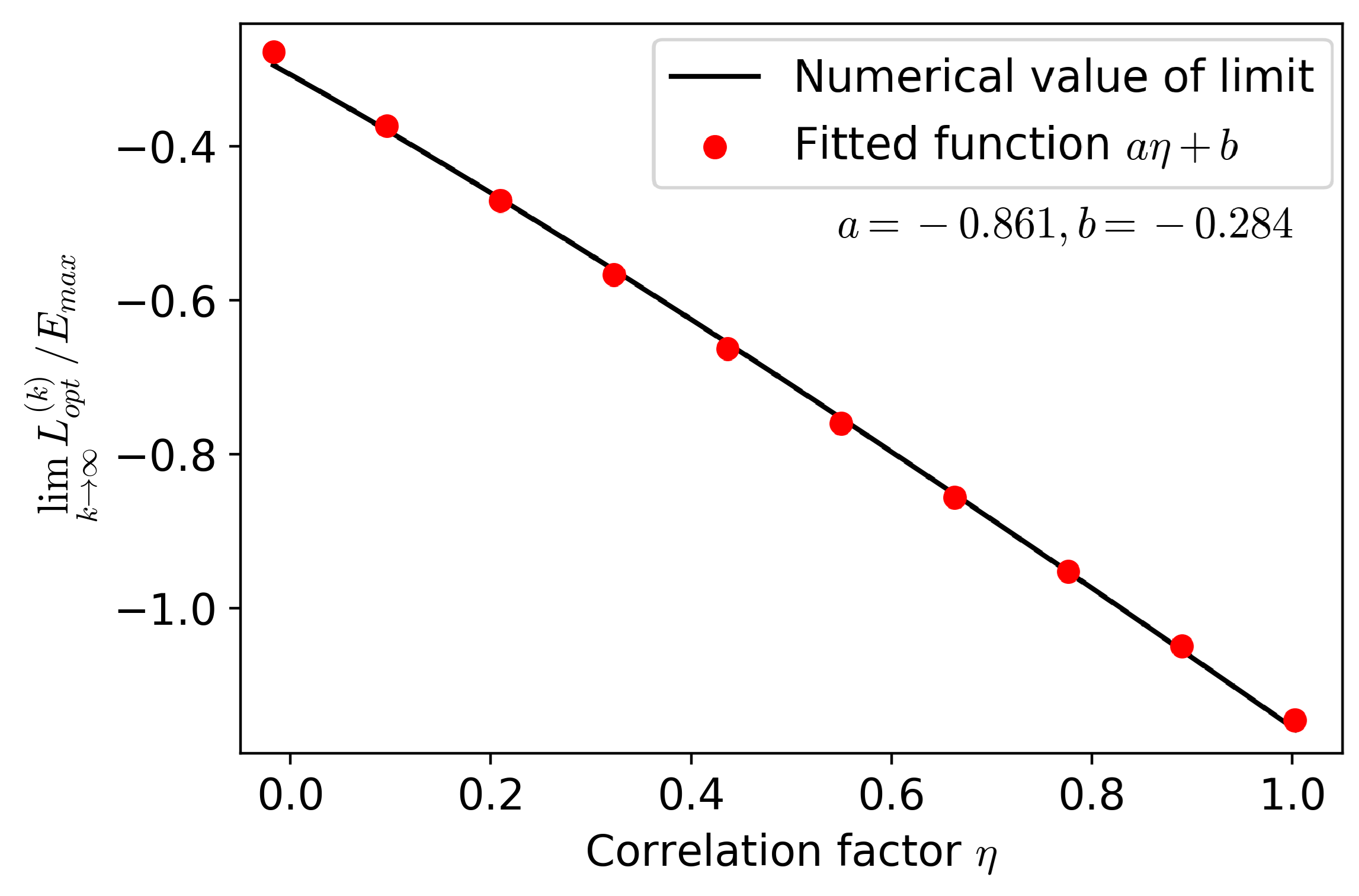}
	\caption{Asymptotic Bell violation for uniformly distributed angles $\vec{\theta}_\mu$ in $[\gamma,\pi-\gamma]$ for $\gamma\approx0.22\pi$.}
	\label{fig:L_infinity}
\end{figure}

Figure~\ref{fig:asymptotic} presents the minimal Bell correlator as a function of the number of measurement settings $k$ for $\eta = 0.2$. 
The mixing term systematically reduces the Bell correlator values across all $k$. In the $k\to \infty$ limit, the  Fig.~\ref{fig:L_infinity} reveals a pronounced enhancement of Bell correlations when additional correlations between symmetric and anti-symmetric subsystems are introduced.

\section{Mixed states for spin-1 BECs}
\label{app:mixedstates}

To explore the noise effect, we consider the mixed states of the form:
\begin{equation}
	\hat{\rho} = p\, \hat{\rho}_S + (1-p)\, \hat{\rho}_\perp,
\end{equation}
where $\hat{\rho}_S$ represents a squeezed Bose-Einstein condensate (BEC) optimized for Bell correlations $L(\hat{\rho}_S)=L^{(k)}_{{\rm opt}}$, and admixture $\hat{\rho}_\perp$ imposes the lower bound on the witness of Bell correlations due to a lost of correlations in the system.
We consider the two different admixture states:
$(i)$ the spin-coherent state $\hat{\rho}_\perp= |\psi_0\rangle \langle \psi_0|$,
and $(ii)$ the maximally mixed spin state $\hat{\rho}_\perp= \mathbb{I}/\mathcal{N}$ where $\mathbb{I}$ is identity matrix and $\mathcal{N}$ normalization factor.

We first study the spin-coherent admixture, for which the corresponding density matrix is given by:\begin{equation}
	\hat{\rho} = p\, \hat{\rho}_S + (1-p)\, |\psi_0\rangle \langle \psi_0|.
\end{equation}
In this case, the non-zero expectation values of the collective operators and the second moment are given by:
\begin{align}
	\langle \hat{J}_{z,1}^{(\sigma)}\rangle &= p \langle \hat{J}^{(\sigma)}_{z, 1}\rangle_S + (1-p) N/2 \\
	\langle (\hat{J}^{(\sigma)}_{\rm{min},1})  ^2\rangle &= p \langle (\hat{J}^{(\sigma)}_{\rm{min},1})^2 \rangle_S 
	+ (1-p) N/4,
\end{align}
with the initial conditions
$\langle \hat{J}^{(\sigma)}_{z, 1} \rangle_S(t=0)=N/2$ and $\langle (\hat{J}^{(\sigma)}_{\rm{min},1} )^2\rangle_S(t=0) =N/4$ and where the lower index $S$ refer to expectation value over the state $\hat{\rho}_S$.
Therefore, the value of $v$ and $\xi^2$ in (\ref{eq:violation for k=3}) are determined by
\begin{align}
	v&=p v_S +(1-p)/2\\
	4 \xi^2 v^2 & =  p (4 v_S^2 \xi_{S}^2) +1-p.
\end{align}
Note that in the thermodynamic limit
$v_S\to 1/2$ and $v\to 1/2$ (unchanged by $p$) while $4 \xi^2 v^2 \to 1-p$.
The lowest value of the latter is non-zero even for $N\to \infty$ and $\xi_S^2 \to 0$. The admixture introduces a lower limit on the value of best squeezing, such that
$4 \xi^2 v^2 \ge 1-p$.
The critical (minimal) value of $p$ for the Bell correlations witnessing is $p_c =1/2$.

\begin{figure}[t]
	\centering
	\includegraphics[width=0.95\linewidth]{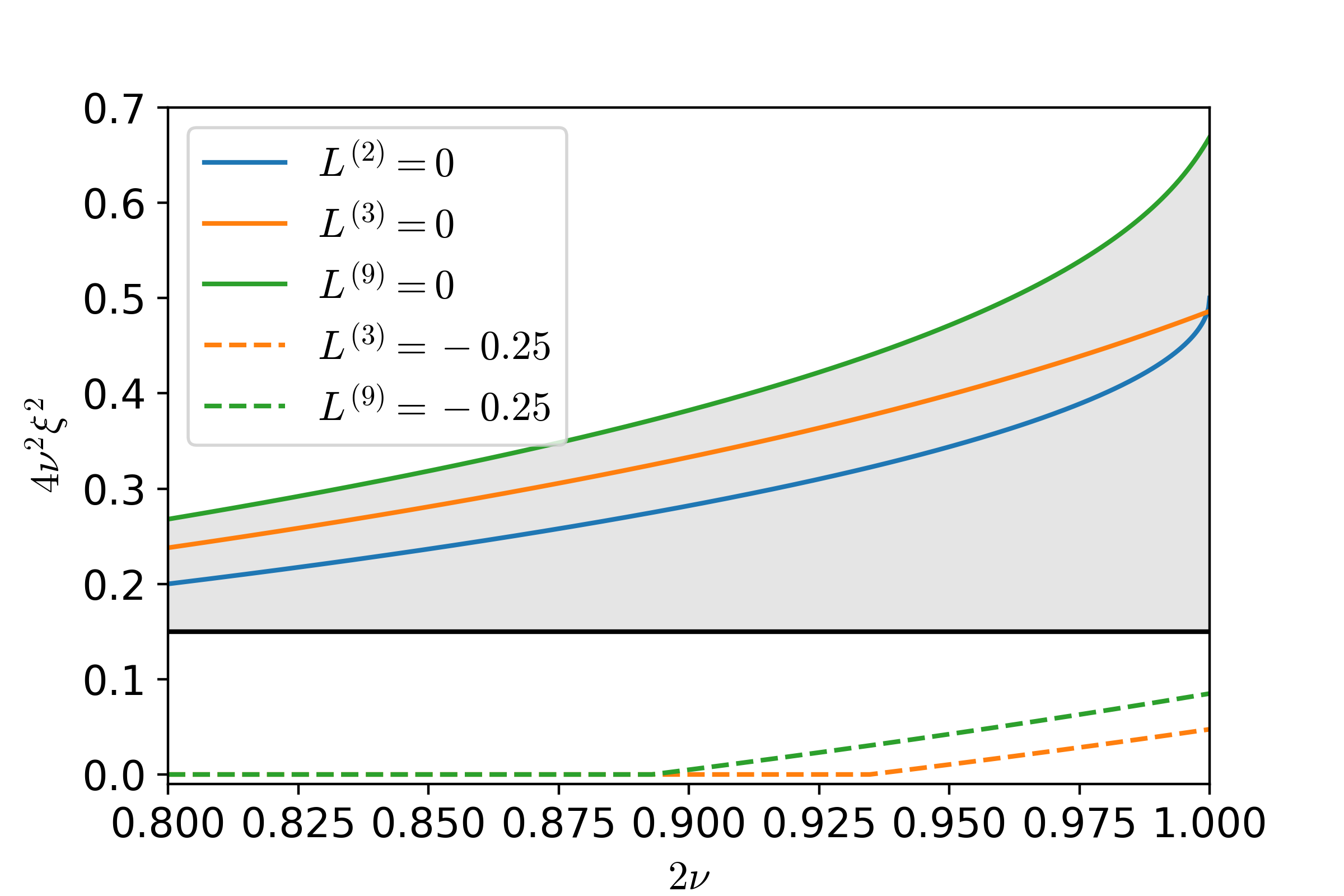}
	\caption{The critical lines for the correlator $L^{(k)}_{{\rm opt}}  =0 $ are presented with solid lines for $k_\mu=2$ (blue), $k_\mu=3$ (orange) and $k_\mu=9$ (green). The violation of Bell inequality is possible in the shaded region. The dashed lines correspond to the case $L^{(k)}_{{\rm opt}}  =-0.25 $, which is the optimal violation for $k_\mu =2$. 
		The noise effect for $p=0.85$, as discussed in the text, reduce the range of parameters $v $ and $\xi^2 $ for Bell correlation violation, marked here with the shaded area. }
	\label{fig:figappB}
\end{figure}

Next, we consider the case of a maximally mixed admixture, where the corresponding density matrix is given by:\begin{equation}
	\label{eq:maxmixstate}
	\hat{\rho} = p\, \hat{\rho}_S + (1-p)\, \mathbb{I}/\mathcal{N},
\end{equation}
with the identity operator in the SU(2) subspace
\begin{align}
	\mathbb{I} &= \sum_{S=0}^{S_{\rm max}-1}  \sum_{m=-S}^S|S, m\rangle \langle S, m|.
\end{align}
Here, the density operator $\hat{\rho}_S$ resides in the Hilbert subspace characterized by the maximal spin quantum number $S=S_{\rm max}=N/2$.
The resulting expectation values of the collective operators' read
\begin{align}
	&\langle \hat{J}_{\vec{n}}\rangle = p \langle \hat{J}_{\vec{n}}\rangle_{S=S_{\rm max} }
	+ (1-p)\sum_S \frac{\langle \hat{J}_{\vec{n}}\rangle_{S} }{\mathcal{N}}
	\label{eqapp:B9} \\
	&\langle (\hat{J}_{\text{min},1}^{(\sigma)})^2\rangle = p \langle (\hat{J}_{\text{min},1}^{(\sigma)})^2 \rangle_{S=S_{\rm max} }
	+ (1-p)\sum_S \frac{\langle (\hat{J}_{\text{min},1}^{(\sigma)})^2 \rangle_{S}}{\mathcal{N}}.
\end{align}
To estimate the corresponding averages, we adopt the following approximations:
$\sum_S \langle \hat{J}_{\vec{n}}\rangle_{S}/\mathcal{N}  = \frac{1}{2}\langle \hat{J}_{\vec{n}}\rangle_{S=S_{\rm max} }$
and
$\sum_S \langle (\hat{J}_{\text{min},1}^{(\sigma)})^2 \rangle_{S}/\mathcal{N} = \frac{1}{2} \langle (\hat{J}_{\text{min},1}^{(\sigma)})^2 \rangle_{S=S_{\rm max} }$.
Therefore, in (\ref{eq:violation for k=3}) we would have
\begin{equation}
	v=p v_S +(1-p) \left(\frac{1}{2} - \frac{1}{N}\right)\frac{1}{2} \approx v_S(p+1)/2
\end{equation}
because $\sum_S \langle \hat{J}_{\vec{n}}\rangle_{S}/\mathcal{N}\approx (N/2-1)/2\approx N v_S/2$.
In addition, we have
\begin{equation}
	4 \xi^2 v^2 \approx (4 v_S^2 \xi_{S}^2) (1+p)/2
\end{equation}
when $\sum_S \langle (\hat{J}^{(\sigma)}_{\text{min},1})^2 \rangle_{S} N/(4 \mathcal{N} ) \approx (4 v_S^2 \xi_{S}^2)/2$.
In the thermodynamic limit, these reduce to
$v_S\to 1/2$ while $v\to (p+1)/4$ and $4 \xi^2 v^2 \to (1-p)/2$. 
From these relations, we derive a critical value $p_c =1/2$
for the Bell correlation witness.

Figure~\ref{fig:figappB} depicts critical lines corresponding to the parameters $4 v^2 \xi^2$ and $v$ at which the Bell correlator equals zero, $L^{(k)}_{{\rm opt}}  =0 $. The case of two measurement settings $k_\mu=2$ as defined in ~\cite{PhysRevLett.119.170403, doi:10.1126/science.1247715, doi:10.1126/science.aad8665}, is represented by a solid blue line, while the case of $k_\mu=3$ and $k_\mu=9$ are denoted by solid orange and red lines, respectively. The enhancement over the two measurement settings can be observed when $L^{(k)}_{{\rm opt}}  < -0.25 $, which is marked by the dashed lines in Fig.~\ref{fig:figappB}.
The noise considered here imposes limitations on the range of parameters $v$ and $\xi^2$ where the Bell correlations can be observed. These boundaries, indicated by solid and dotted black lines in Fig.~\ref{fig:figappB} for $p=0.85$, delineate the region where correlations remain detectable under the considered noise model.

\bibliography{bibliography}

\end{document}